\documentclass[a4paper,aps,prd,preprintnumbers,showpacs,superscriptaddress,nofootinbib,amsmath,amssymb,twocolumn]{revtex4-2}
\usepackage{tabularray}
\usepackage{graphicx}
\usepackage{bm}
\usepackage{placeins}
\usepackage{xspace}
\usepackage{bm,amsmath,amssymb,amsfonts,gensymb,bbold}
\usepackage{comment}
\usepackage{hyperref}
\hypersetup{
    colorlinks=true,
    linkcolor=blue,
    citecolor=blue,
    filecolor=blue,      
    urlcolor=blue,
}
\usepackage{nicefrac}
\usepackage{cleveref}

\newcommand{\eq}[1]{\begin{align} #1 \end{align}}

\usepackage{color}

\usepackage[english]{babel}
\usepackage{verbatim}
\usepackage{makecell}
\DeclareMathOperator{\sech}{sech}
\usepackage{ amssymb }
\begin{document}

\title{Quasinormal modes and gray-body factors of regular black holes in asymptotically safe gravity}
\author{Oleksandr Stashko}\email{alexander.stashko@gmail.com}
%\affiliation{277 Nassau street, Princeton, NJ, 08540, USA}
\affiliation{Princeton University, Princeton, NJ, 08544, USA}
\begin{abstract}
Recently, Bonanno et al. (2024) presented an explicit metric describing the exterior of a collapsing dust ball within the framework of asymptotically safe gravity. Based on this metric, we investigate in detail the properties of the quasinormal mode (QNM) spectra for test massless scalar, vector, and Dirac fields. We find accurate values for the fundamental QNM frequency and the first overtones, demonstrating the appearance of a peculiar spiral-like behavior of the overtones compared to the fundamental modes. Additionally, we study black hole shadows and show that they align with EHT observations. Finally, we study the grey-body factors and provide explicit lower bound estimates.
\end{abstract}

\maketitle

\section{Introduction}
According to Penrose's singularity theorem \cite{PhysRevLett.14.57,Hawking:1970zqf,Senovilla_1998}, singularities are classically inevitable in General Relativity. Their presence leads to a loss of predictability and the breakdown of standard physical laws. The cosmic censorship conjecture \cite{Penrose:1969pc} states that such singularities should be hidden beneath the event horizon of a black hole (BH). Consequently, the interest in regular (non-singular) black hole models and the exploration of ways to resolve singularities has remained a central focus of gravitational physics for several decades.

The first regular black hole solution was proposed by Bardeen \cite{Bardeen}, who considered the collapse of charged matter with an AdS matter core inside the black hole instead of a singularity. Since then, numerous regular black hole solutions have been obtained using different matter sources \cite{PhysRevLett.80.5056,1992GReGr..24..235D,PhysRevLett.96.031103,2006Bronnikov,Maeda_2022,2002PhRvD..65f4039B,Fan_2016,PhysRevD.103.084052,2023arXiv230712357G,Ovalle_2023}  or within modified gravity theories \cite{Bueno:2024dgm,2024CQGra..41e5012J,Carballo_Rubio_2022,2024PhRvD.109b4032B,2006GReGr..38..885B,Mazza_2023,2023EPJC...83..475J}, etc. Similar attempts have been made by incorporating quantum corrections, at least at a phenomenological level \cite{PhysRevD.76.104030,2010PhRvD..82j4035M,2023JHEP...05..118C,2023PhRvD.108f1502F,2019PhLB..79734888N,Knorr_2022,Platania_2019, PhysRevD.79.084023,2010IJTP...49.1649M,Koch_2013,Bonanno_2000,PhysRevLett.130.101501,Bonanno_etal2024PhRvL,PhysRevD.106.046006,PhysRevD.108.044051,Bonanno_etal2024PhRvL} or by regularization of singularities \cite{2020PhRvD.101h4047C}. 

Considerable attention has also been given to the concept of asymptotic safety \cite{Bonanno_2000,Bonanno_etal2024PhRvL,Eichhorn2012PhRvD, Niedermaier_2007,Platania_2023,PhysRevD.109.024045,Borissova_2023, pawlowski2023effectiveactionblackhole}.
Recently, Bonanno et al. \cite{Bonanno_etal2024PhRvL} proposed a new model of a regular black hole based on the ideas of Markov and Mukhanov, utilizing the concept of gravity’s antiscreening behavior in ultra-Planckian energy domains. They presented an explicit metric describing the exterior of the collapsing dust ball within the framework of asymptotically safe gravity.

It is assumed that quantum corrections, which blur the singularity inside a black hole, act on Planck scales and are unlikely to produce noticeable astrophysical effects since, far from the horizon, the black hole geometry tends to be Schwarzschild.  Therefore, a natural question arises how these quantum corrections can be potentially measurable.

Recent detection of gravitational wave bursts from  stellar-mass BH collisions  \cite{LIGO+VIRGO2016,LIGO-VIRGO-KAGRA2023}, and the observations of BH shadows \cite{Akiyama2019,EHT2022} provide an opportunity for testing subtle gravitational effects that characterize compact relativistic objects and have opened up possibilities for new fundamental physics. 
\cite{LISA2022LRR}. In this context, numerous studies have investigated BH shadows \cite{Abdujabbarov_2016,Ling_2022,Ghosh_2020,PhysRevD.107.124003,universe5070163,Yang_2023,2022ApJ...939...77W} and the quasi-normal modes (QNM) \cite{Flachi_2013,2023PhRvD.107j4050K,PhysRevD.109.044044,2021Univ....7..418S,2022JCAP...10..091K,2021AnPhy.43468603M,2023CQGra..40s5024M,2024JCAP...01..020F,PhysRevD.108.104054,PhysRevD.109.104005,zhang2024quasinormal,Ghosh_2023,2024arXiv240407589P} of various regular BH.

In the present paper, we study in detail the behavior of test massless scalar, vector, and Dirac $(s=1/2)$ fields. We demonstrate that while the fundamental mode is weakly sensitive to the scale parameter $\xi$,  the behavior of the overtones exhibits a peculiar feature related to the spiral-like form of the QNM trajectories as a function of $\xi$.

This paper is organized as follows. In Sec. \ref{sec:BR}, we provide the basic information about the black hole space-time under consideration. The Sec. \ref{sec:wavelike}  is devoted to the master wave-like equations of the test fields and their effective potentials. In Sec. \ref{sec:methods}, we briefly explain the Chebyshev pseudospectral method used for calculations of the quasinormal spectra and properties, which we use to calculate quasinormal modes and discuss in detail the properties of the quasinormal spectra for different values of parameters. In Sec. \ref{sec:Eikonal}, we examine the properties of black hole shadows and quasinormal modes in the eikonal approximation. In Sec. \ref{sec:scattering}, we analyze the effect of quantum corrections on the grey-body factor. The paper concludes with a short summary in Sec. \ref{sec:conclusions}. In appendix, we provide tables with accurate values of fundamental quasinormal modes and the first overtones.

\section{Basic relations}
\label{sec:BR}
Recently, Bonnano et al. \cite{Bonanno_etal2024PhRvL} extended the ideas of Markov and
Mukhanov \cite{1985NCimB..86...97M} and considered the appearance of regular black holes during the gravitational collapse. They deal with the following action:
\begin{equation}
\label{effS}
S = \frac{1}{16 \pi G_N} \int d^4 x \sqrt{-g} \left[R + 2  \chi(\epsilon)  \mathcal{L}\right],
\end{equation}
where $\chi = \chi(\epsilon)$ is some coupling\footnote{ $\chi(\epsilon=0) = 8 \pi G_N$} and $\mathcal{L}$ is the matter Lagrangian, respectively.

As a result,  the total variation of the action (\ref{effS}) yields the following equations
\eq{
\label{effectiveEQ}
R_{\mu\nu}- \frac{1}{2}g_{\mu\nu}R = 8\pi G(\epsilon)
T_{\mu\nu}-\Lambda(\epsilon) g_{\mu\nu},
}
where $T_{\mu\nu}= [\epsilon+p(\epsilon)] u_\mu u_\nu+p g_{\mu\nu}$  is the energy-momentum tensor of ideal fluid and $G(\epsilon)$, and $\Lambda(\epsilon)$ denote the effective gravitational and cosmological constants, defined as
\begin{equation}
\label{efeg}
8\pi G(\epsilon)=\frac{\partial (\chi \epsilon)}{\partial \epsilon}, \quad \Lambda(\epsilon)=-\frac{\partial \chi}{\partial \epsilon} \epsilon^2,
\end{equation}

The behavior of $G(\epsilon)$ as a function of the energy scale is governed by a renormalization group trajectory close to the ultraviolet fixed point of the Asymptotic Safety program \cite{ PhysRevD.57.971, Bonanno_2020,10.21468/SciPostPhys.12.1.001,PhysRevD.62.043008} and was taken as
\begin{equation}
\label{G}
G(\epsilon)=\frac{G_N}{1+\xi \epsilon},
\end{equation}
where $G_N$ is the ordinary gravitational constant and $\xi$ is a scale parameter. The exact value of $\xi$ is unknown  and should be constrained  from the observations \cite{Bonanno_etal2024PhRvL}.

As a result of the gravitational collapse of dust ($p=0$), they derived  the following metric of the static exterior space-time \cite{Bonanno_etal2024PhRvL}
\eq{\label{eq:metric}
ds^2 = f(r) d t^2- \frac{dr^2}{f(r)} - r^2 (\sin^2 \theta d\phi^2+d\theta^2),
}
with
\begin{equation}\label{eq:f(r)}
f(r) = 1-\frac{r^2}{3 \xi} \ln \left(1 + \frac{6 M \xi }{r^3}\right)\,,
\end{equation}
where $M$ is the mass of the configuration\footnote{$G_N=c=1$}.

There are  critical values 
\begin{equation}
\xi_{\rm cr}=\frac{2}{3}y^2(3+2y)M^2,~~r_h^{\rm ext}=-2y M,~~y=W_0\left(-\frac{3}{2e^{3/2}}\right), 
\end{equation}
where $W(x)$ is the Lambert function and numerical values are  $\xi_{\rm cr}\simeq 0.4565M^2$ and $r_h^{\rm ext}\simeq 1.2516M $

(1) For $\xi \in (0, \xi_{\rm cr})$, there are two roots of $f(r)$ that correspond to the inner horizon $r_h^{\rm in} \in (0, r_h^{\rm ext})$ and the outer horizon $r_h \in (r_h^{\rm ext}, 2M)$. For small values of $\xi$, the outer horizon $r_h$ can be approximated as
\eq{r_h = 2M - \frac{3\xi}{4M} - \frac{15\xi^2}{32M^3} + O(\xi^3).}

(2) For $\xi = \xi_{\rm cr}$, the inner and outer horizons coincide into a single horizon at $r_h = r_h^{\rm ext}$, and the solution corresponds to an extremal black hole.

(3) For $\xi>\xi_{\rm cr}$, we have a horizonless configuration.

In the classical limit $\xi\to{0}$, we obtain the Schwarzschild solution.

\section{Master equations}\label{sec:wavelike}
We are going to study the behavior of massless test scalar ($\Phi$), massless vector ($A_\mu$), and Dirac ($\Upsilon$) fields on the background of (\ref{eq:metric}). The general relativistic equations for these matter fields can be written in the following form 
\eq{\label{eq:kg}
\frac{1}{\sqrt{-g}}\partial_\mu \left(\sqrt{-g}g^{\mu \nu}\partial_\nu\Phi\right)&=0,\\
\label{eq:em}
\frac{1}{\sqrt{-g}}\partial_{\mu} \left(F_{\rho\sigma}g^{\rho \nu}g^{\sigma \mu}\sqrt{-g}\right)&=0\,,\\
\label{eq:dirac}
\gamma^{\alpha} \left( \frac{\partial}{\partial x^{\alpha}} - \Gamma_{\alpha} \right) \Upsilon&=0,
}
where $F_{\mu\nu}=\partial_\mu A_\nu-\partial_\nu A_\mu$ is the vector field tensor; $\gamma^{\alpha}$  and $\Gamma_{\alpha}$ are noncommutative gamma matrices and  spin connections in the tetrad formalism \cite{RevModPhys.29.465}, respectively.

Equations (\ref{eq:kg}), (\ref{eq:em}), and (\ref{eq:dirac}) can be reduced to one master equation with respect to one scalar function $\Psi$
\cite{Konoplya_2011,Arbey_2021}
\eq{\label{eq:masterEq}
\dfrac{\partial^2 \Psi}{\partial t^2}-\dfrac{\partial^2 \Psi}{\partial r^{*}{}^2}+ V^{(i)}_{\rm eff}(r)\Psi=0,
}
where $r=r(r^*)$, is the ``tortoise coordinate''  defined as
$dr^*={dr}/f(r)$.

In the case of the massless scalar field ($i=s$) the effective potential is
\eq{\label{potentialScalar}
V^{(s)}_{\rm eff}(r)=f(r)\left( \frac{l(l+1)}{r^2}+\frac{1}{r}\frac{d f(r)}{dr}\right),
}
For the massless vector field ($i=v$), the effective potential is
 \eq{
V^{(v)}_{\rm eff}(r)=f(r)\left( \frac{l(l+1)}{r^2}\right),
}
And for the massless Dirac field ($i=d$), we have 
\eq{
     V^{(d,\pm)}_{\rm eff}(r) = f(r)\frac{l+1}{r}\left(\frac{l+1}{r}\mp \sqrt{f(r)} \pm \frac{d}{dr}\sqrt{f(r)} \right)\,,
}
where  $V^{(d,\pm)}_{\rm eff}$ correspond to the spin up and down fermions.  It was shown that potentials $V^{(d,+)}_{\rm eff}$ and $V^{(d,-)}_{\rm eff}$  are isospectral, and both can be transformed into each other with the help of the Darboux transformation \cite{Zinhailo_2019,Konoplya_2011}. Further, we restrict ourselves with the $V^{(d)}_{\rm eff}(r)\equiv V^{(d,+)}_{\rm eff}(r)$ case.
  
\section{Quasinormal modes}\label{sec:methods}
\subsection{Chebyshev pseudo-spectral method}
Quasinormal frequencies are the eigenvalues $\omega$ of the master equation (\ref{eq:masterEq}), which  correspond to in-going waves at the black hole horizon and outgoing waves at spatial infinity. They satisfy the following boundary conditions
\eq{
\Psi\sim  e^{-i\omega(r^*+t)},~ r \to r_{\rm h},\,\,\,\Psi\sim e^{i\omega(r^*-t)},~ r \to \infty.
}

In order to solve the corresponding  eigenproblem, we use  the Chebyshev pseudo-spectral method \cite{boyd2013chebyshev}. 
For numerical purposes, it is convenient to rewrite (\ref{eq:masterEq}) using the ingoing Eddington-Finkelstein coordinates $(v=t+r^*,r)$  \cite{2017EPJP..132..546J}.
After separation of variable $v$ in the form $\Psi\sim \exp(-i\omega v)\Psi(r)$, the master equation (\ref{eq:masterEq}) in these new coordinates takes the form 
\eq{\label{eq:MasterEqEFcoord}
f(r)\frac{d^2 \Psi}{dr^2}+\left(\frac{df(r)}{dr}-2i\omega\right)\frac{d \Psi}{dr}-\frac{V^{(i)}_{\rm eff}(r)}{f(r)}\Psi=0,
}
with boundary conditions
\eq{
\Psi\sim const,~ r \to r_{\rm h},\quad \Psi\sim e^{2i\omega r^*},~ r \to \infty\,.
}
correspondingly at the black hole horizon and at spatial infinity.

Pseudospectral methods usually work on a finite interval;  that is why we compactify our semi-interval $[r_h,\infty)$ to $[0,1]$ by introducing a new variable \footnote{In the case of the Dirac field, it is more convenient to use the following compactification:  $r=r_h/(1-u^2)$.}
\eq{
r=\frac{r_h}{1-u},
}
As we use the Eddington-Finkelstein coordinates, we have $\Psi \sim const$ at the horizon, and we only need to separate the correct asymptotic solution at spatial infinity. We substitute
\eq{
\Psi(r)=(1-u)^{-4i\omega M}e^{\frac{2i r_h\omega}{1-u}}\Phi(r),
}
 which gives  
\eq{\Phi\sim const,~u\to{0},\quad \Phi\sim const,~u\to{1}.
}
The final equation might suffer from numerical singularities at the boundaries $u=0$ and $u=1$. To avoid them, we can make an additional substitution $\Phi(u) = u(1-u)\psi(u)$, and our boundary conditions can now be written as follows
\eq{
\psi(0)=0,\quad\psi(1)=0.
}
All previous substitutions lead to an equation of the form
\eq{\label{eq:reg}
A_2(u)\psi''(u)+A_1(u)\psi'(u)+A_0(u)\psi(u)=0,
}
where $A_i(u)=A_i(u,\omega,\omega^2)$, $i=0,1,2$. At the next step, we discretize Eq. (\ref{eq:reg}) on the Chebyshev-Lobatto grid, which is defined as
\eq{u_j=\frac{1}{2}\left(1-\cos\left[\frac{\pi j}{N}\right]\right),~~j=0,1...N.}
This results in the matrix equation
\eq{\label{eq:discr_eq}
\left(\tilde{M}_0+\tilde{M}_1
\omega+\tilde{M}_2\,\omega^2\right)\tilde{\psi}=0,
}
where $\tilde{\psi}=[\psi(u_0),...\psi(u_N)]^T$ is the vector of function values at grid points and $\tilde{M}_i$   are matrices of discretized coefficients. 
Following \cite{2017EPJP..132..546J} we rewrite (\ref{eq:discr_eq}) to obtain a linear form of the eigenproblem     
\eq{\label{eq:LinEig_eq}
\left(M_0+M_1\omega\right)\tilde{\psi}_1=\mathbb{0},}
where  
\begin{equation}
M_0=
\begin{pmatrix}
\tilde{M}_0 & \tilde{M}_1 \\
\mathbb{0} & \mathbb{1}
\end{pmatrix},\,
M_1=
\begin{pmatrix}
\mathbb{0} & \tilde{M}_2 \\
-\mathbb{1} & \mathbb{0}
\end{pmatrix},\,
\tilde{\psi}_1=
\begin{pmatrix}
\tilde{\psi}\\
\omega\tilde{\psi}
\end{pmatrix}.
\end{equation}

Then corresponding QNM frequencies can be
determined by solving generalized eigenvalue problem (\ref{eq:LinEig_eq}). The resulting set of the eigenvalues   also contains spurious eigenvalues \cite{boyd2013chebyshev}. To eliminate them we perform  the calculations on two grids of different sizes and select only overlapping values. The typical grid sizes are $N=100-500$, which  is sufficient  to obtain  good accuracy and the first few overtones. The presence of square roots in the effective potential for the Dirac field case might trigger some numerical  instabilities, which affect the convergence of the pseudospectral methods and require higher precision and grid resolution to obtain high overtone values.

\subsection{Quasinormal modes spectra}

\begin{figure}
    \centering
    \includegraphics[width=.49\textwidth]{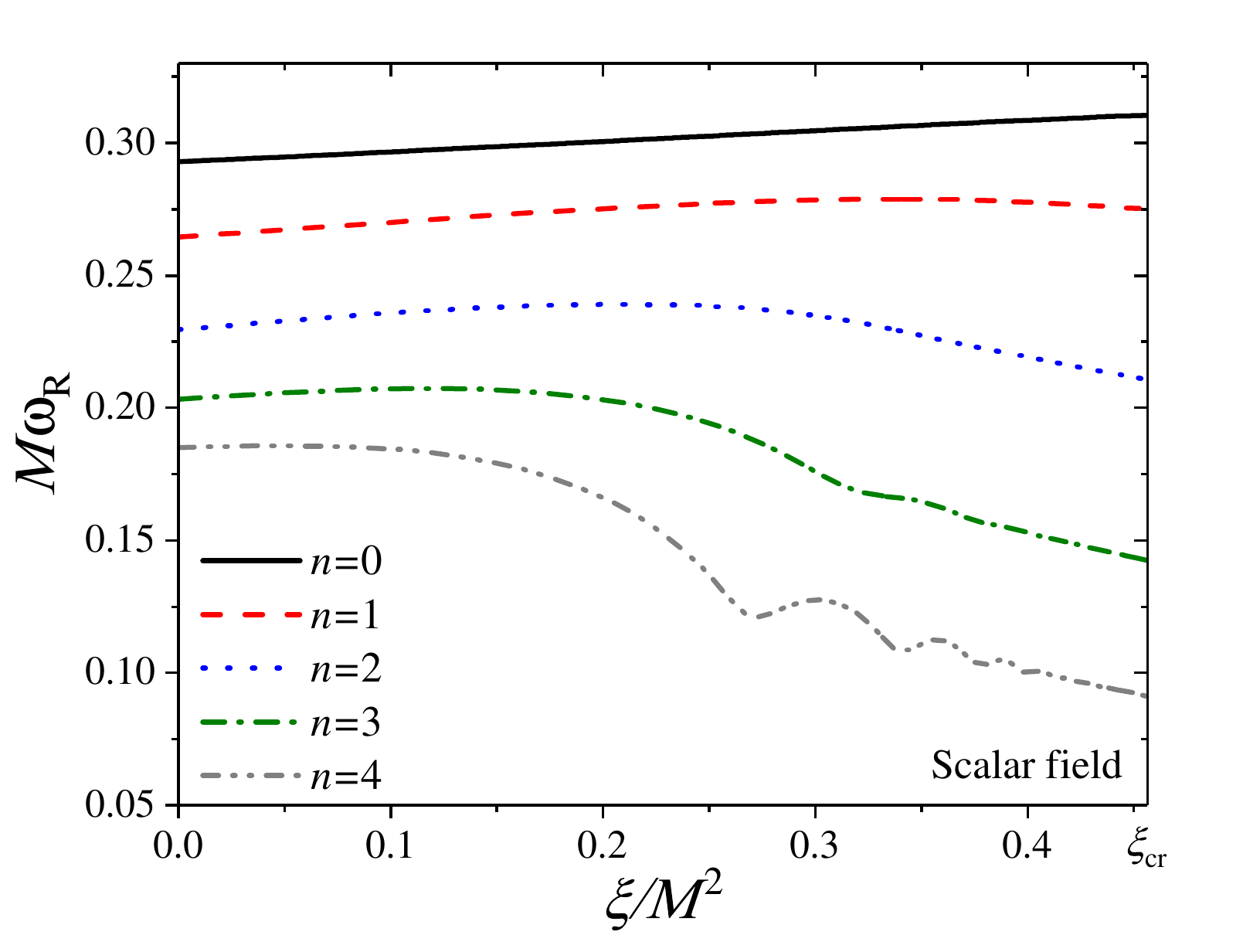}
    \includegraphics[width=.49\textwidth]{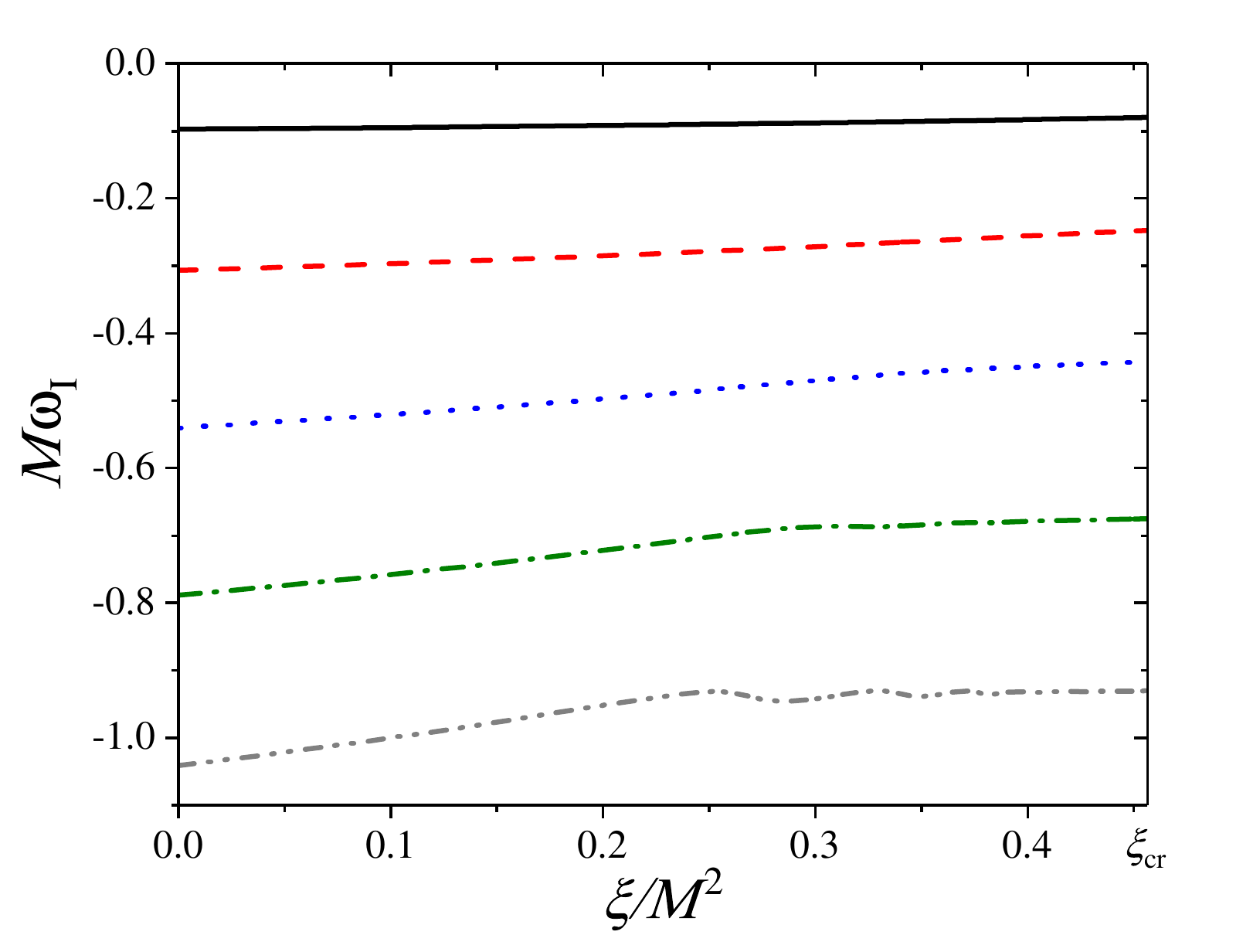}
    \caption{Typical examples of the behavior of the real (top panel) and imaginary (bottom panel) parts of the fundamental quasinormal mode and the first four overtones for the scalar field perturbations with  $l=1$. }
    \label{fig:qnms_scalar_IR}
\end{figure}

\begin{figure}
    \centering
    \includegraphics[width=.49\textwidth]{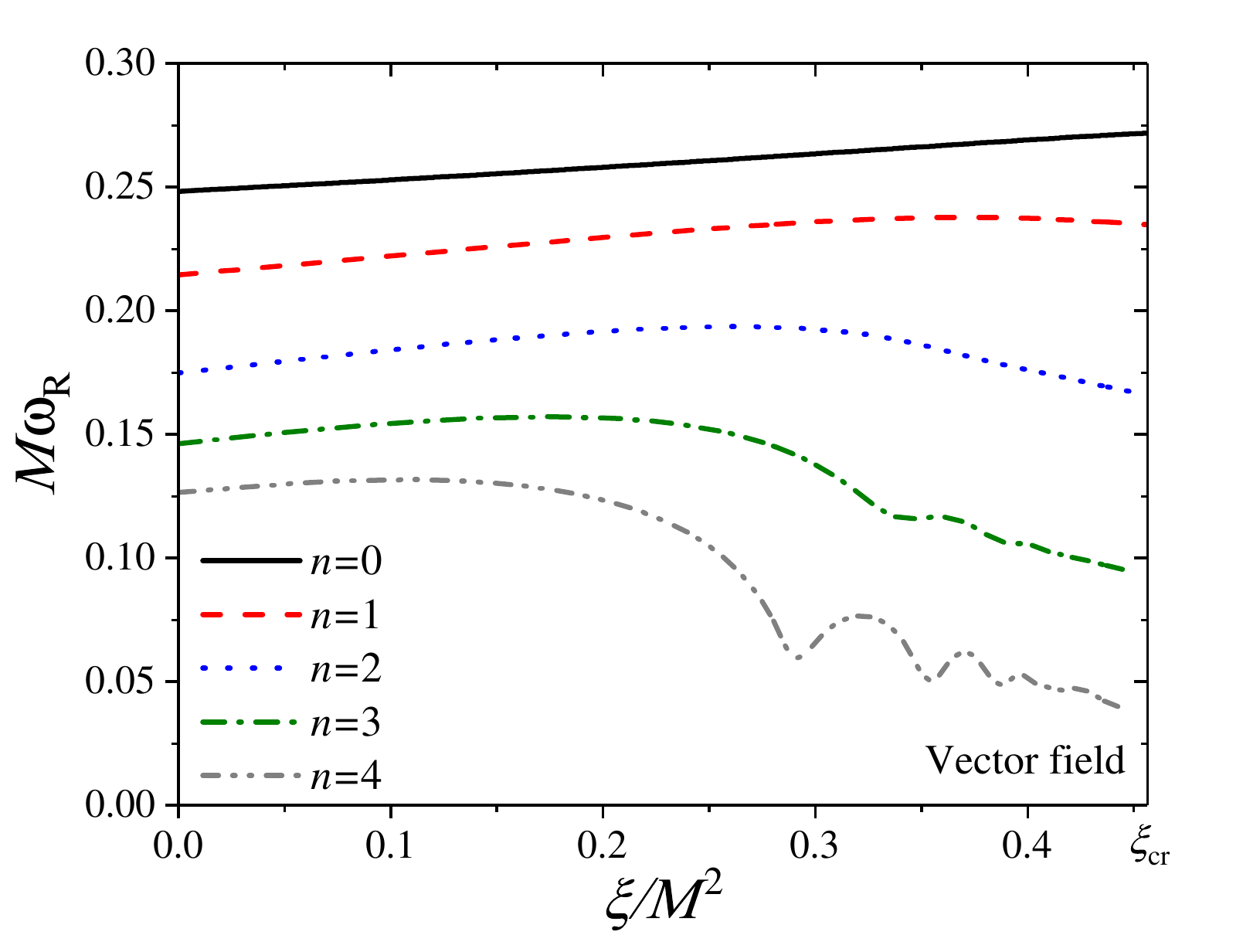}
    \includegraphics[width=.49\textwidth]{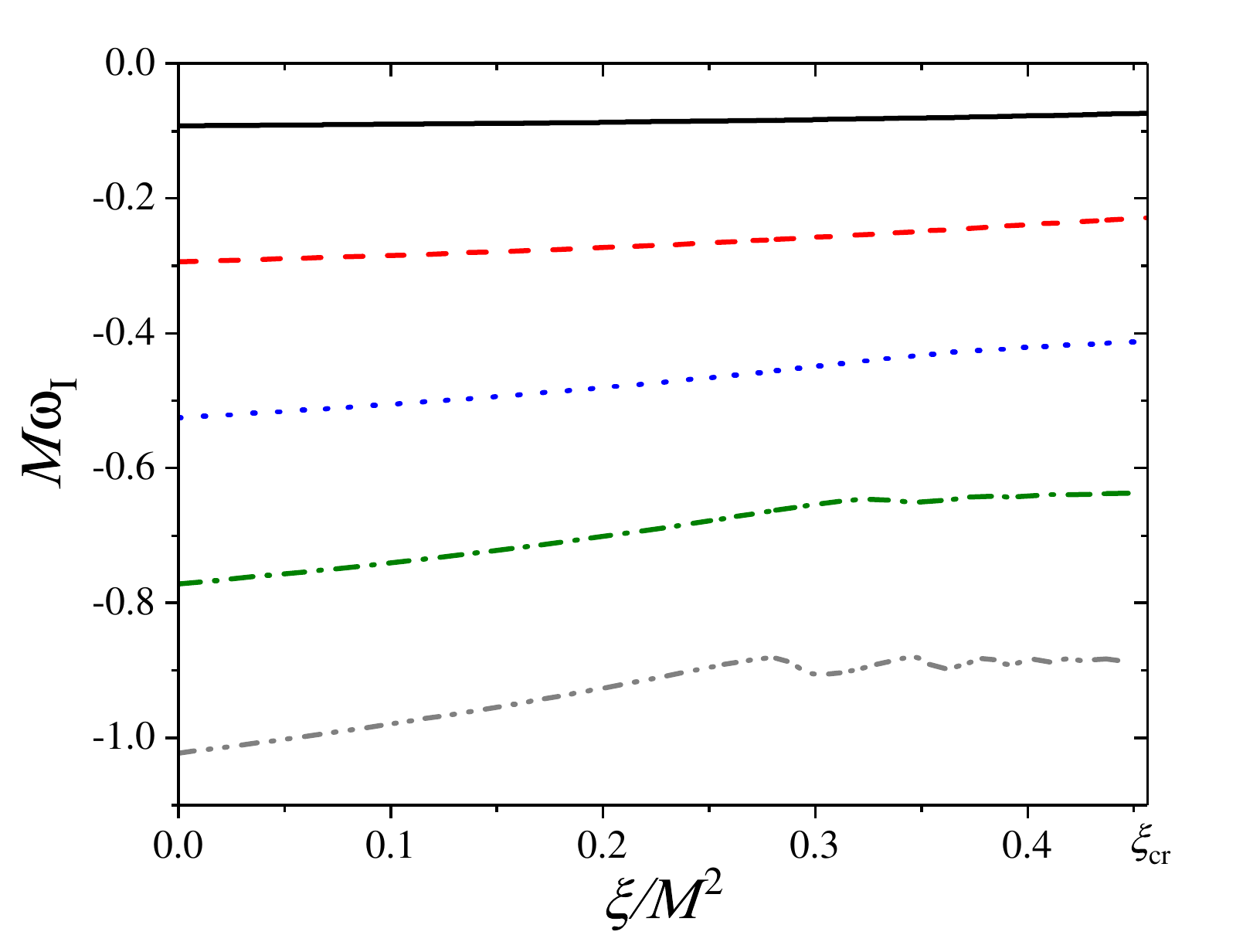}
    \caption{Typical examples of the behavior of the real (top panel) and imaginary (bottom panel) parts of the fundamental quasinormal mode and the first four overtones for the vector field perturbations with  $l=1$. }
    \label{fig:qnms_vector_IR}
\end{figure}

\begin{figure}
    \centering
    \includegraphics[width=.49\textwidth]{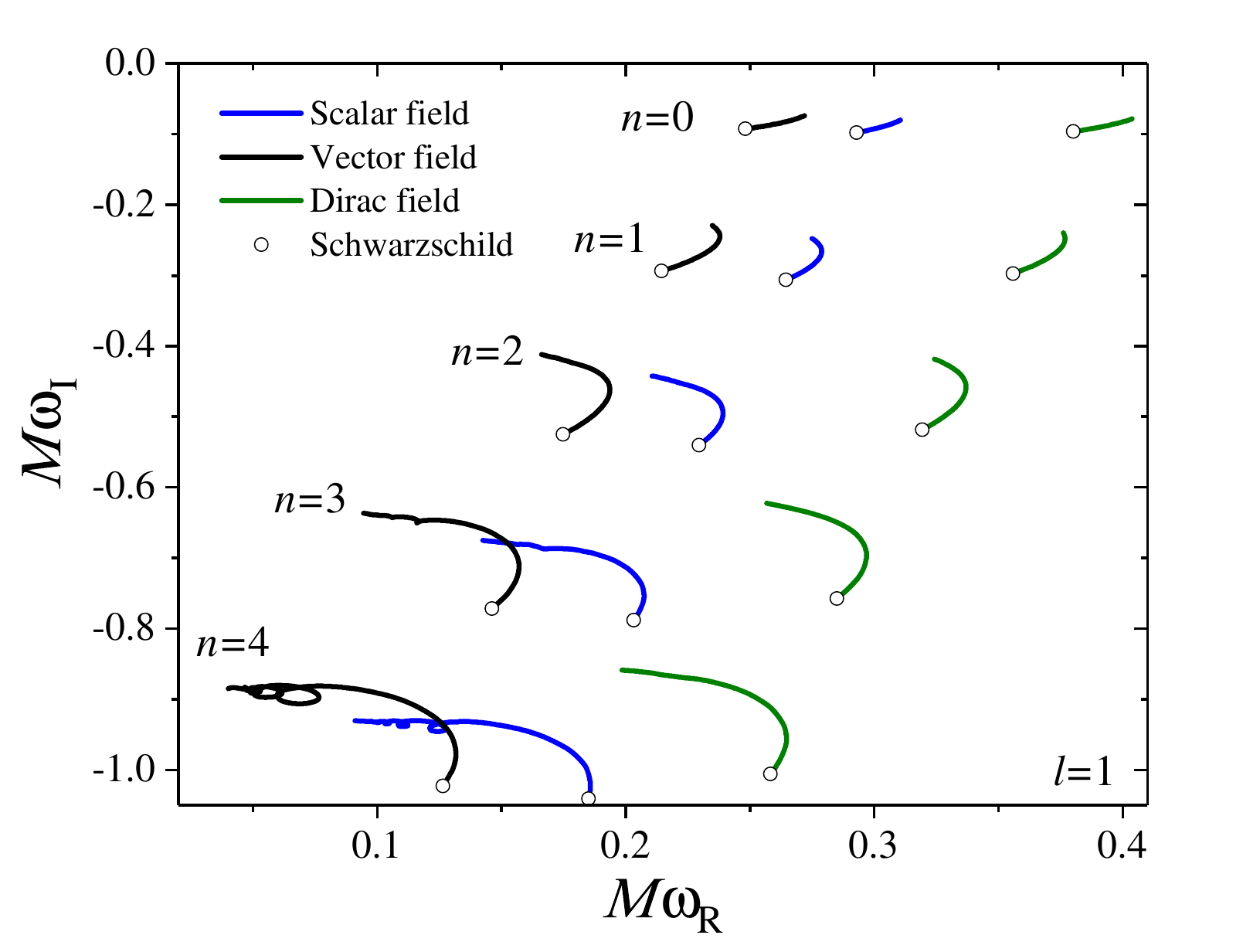}

    \caption{The QNMs trajectories in the complex $\omega$ plane for the fundamental QNMs and the first four overtones for scalar, vector, and Dirac fields with $l=1$.}
    \label{fig:w_plane}
\end{figure}

\begin{figure}
    \centering
    \includegraphics[width=.49\textwidth]{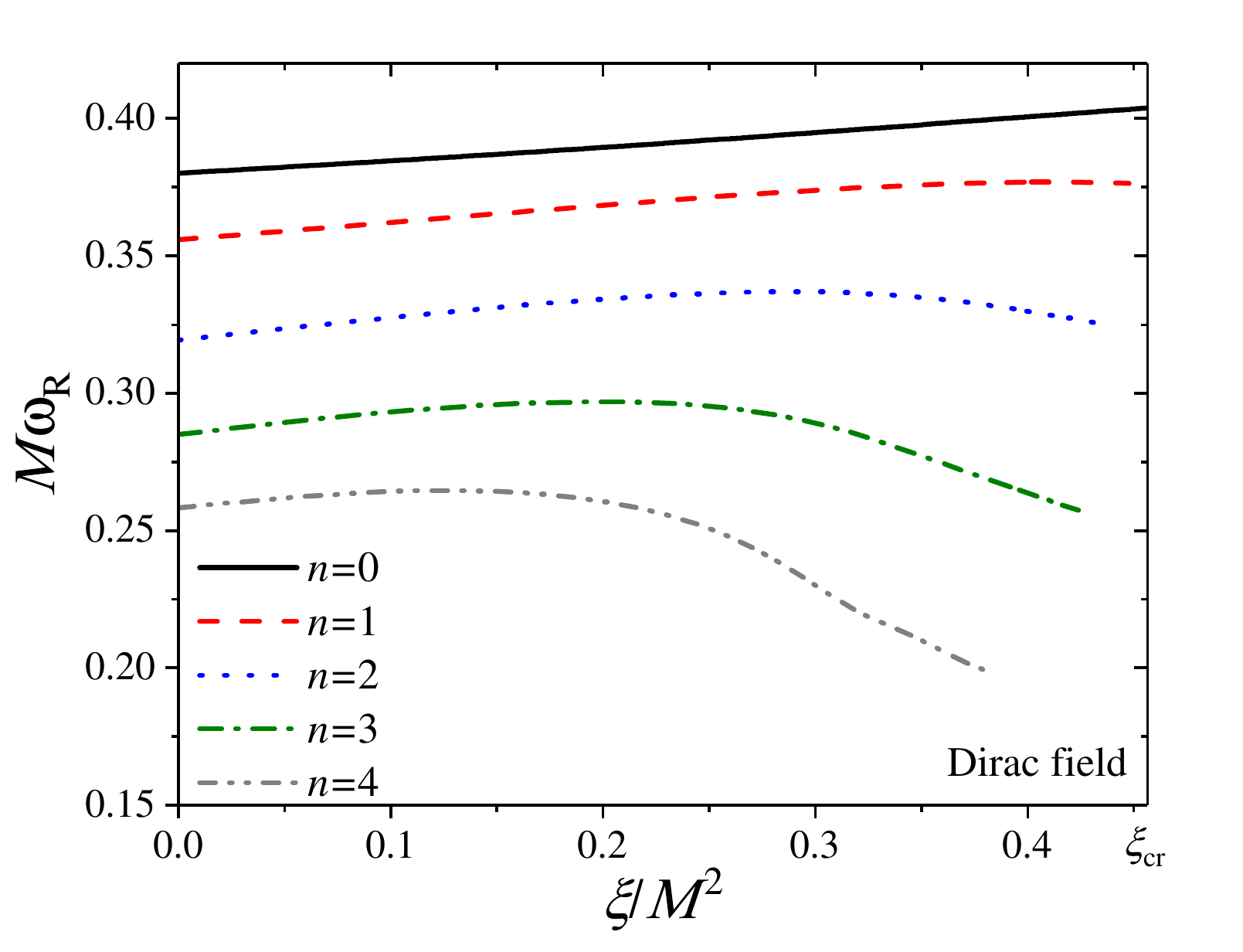}
    \includegraphics[width=.49\textwidth]{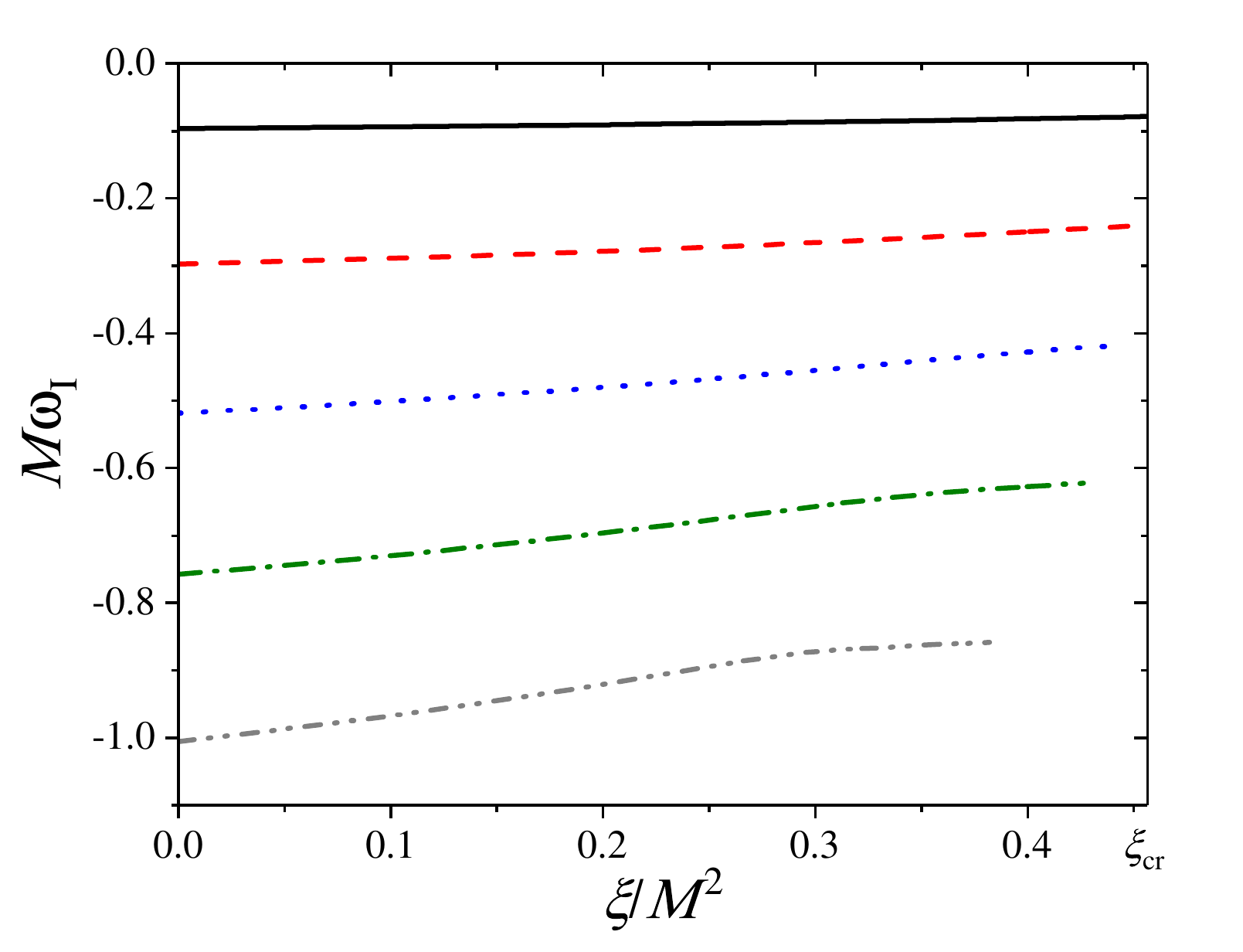}
    \caption{Typical examples of the behavior of the real (top panel) and imaginary (bottom panel) parts of the fundamental quasinormal mode and the first four overtones for the Dirac field perturbations with  $l=1$. The curves stop before the extreme BH case at $\xi=\xi_{\rm cr}$ due to poor convergence of the pseudospectral method.}
    \label{fig:qnms_dirac_IR}
\end{figure}

We present typical dependencies of the fundamental mode and first overtones as functions of $\xi$ for $l=1$ in Figs.  \ref{fig:qnms_scalar_IR}-\ref{fig:qnms_dirac_IR}  for the scalar, vector, and Dirac  massless test fields. The accurate values of QNM frequencies are given in Tabs.  \ref{tab:scalar}-\ref{tab:dirac}.

For the fundamental mode ($n=0$) and $l=1$,  the real $\omega_{\rm R}$ and imaginary $\omega_{\rm I}$ parts of $\omega$ are increasing and decreasing functions of  $\xi$, respectively.  This property of the fundamental mode is inherent to all considered fields. The maximal relative difference between the Schwarzschild black hole ($\xi=0$) and extremal black hole ($\xi=\xi_{\rm cr}$) is less than $10\%$ for the real part and  $21\%$ for the imaginary part of $\omega$.
Starting from the first  overtone ($n=1$) for $\omega_{\rm R}$  and from the second  overtone ($n=2$) for 
$\omega_{\rm I}$, the dependence on 
$\xi$ becomes non-monotonic.

For the higher overtones, such non-monotonicity  leads to a peculiar behaviour of the QNM spectra in the complex $\omega$ plane, relating to the spiral form of the QNM trajectories that spiral toward $\omega(\xi=\xi_{\rm cr})$ as $\xi\to \xi_{\rm cr}$ (see Fig. \ref{fig:w_plane}). Such  spiral-like structure of the QNM spectra has been observed previously for the Reissner-Nordstrom \cite{PhysRevD.68.044027,Jiliang2005}, Kerr \cite{PhysRevD.93.064062}, Hayward \cite{Konoplya_2022} black holes, and in a number of quantum-corrected black holes \cite{2024PhRvD.109b6010F,Moreira_2023,2023arXiv231217639G,2024arXiv240404447G,zhang2024quasinormal}.

\begin{figure}
    \centering
        \includegraphics[width=.49\textwidth]{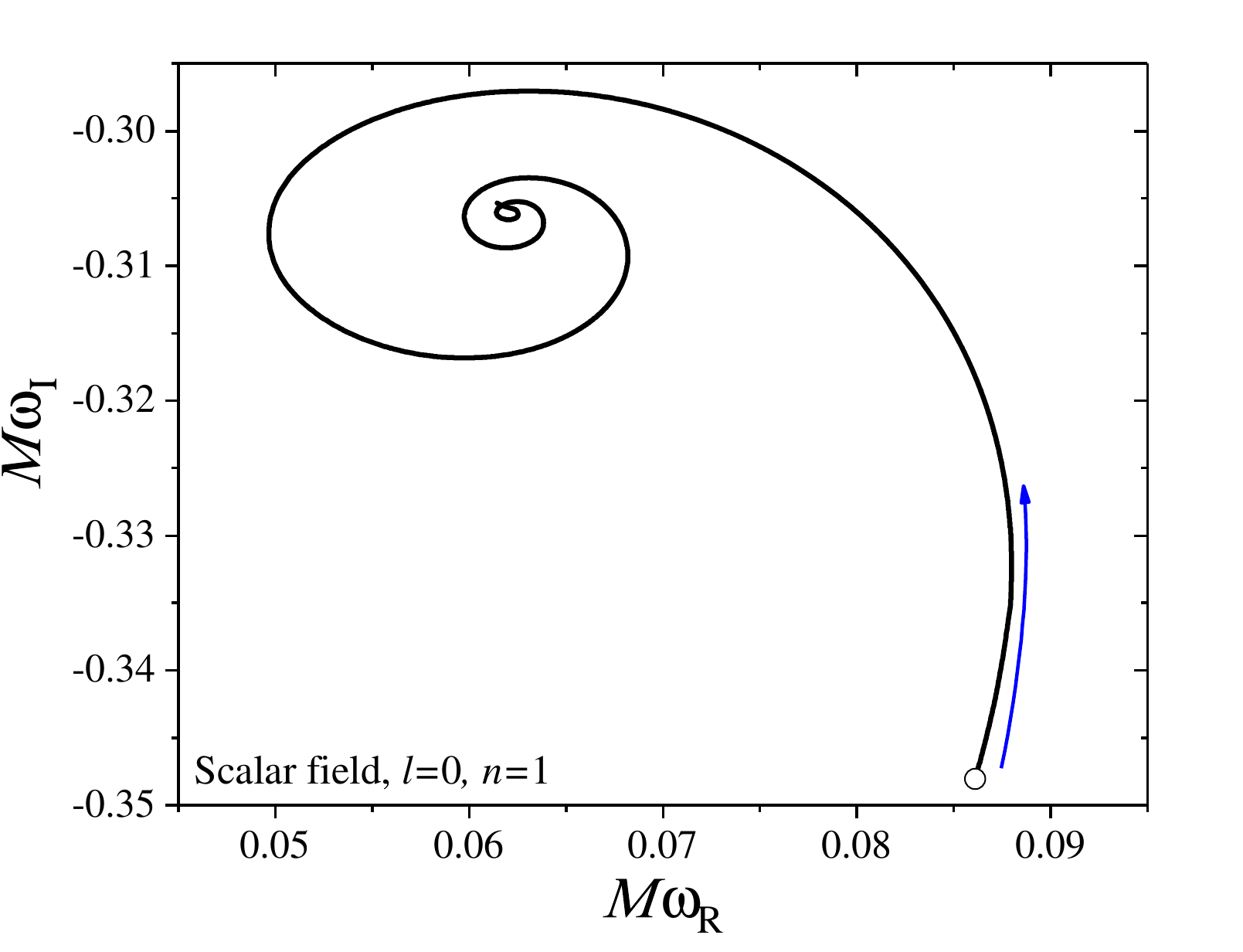}
      \includegraphics[width=.49\textwidth]{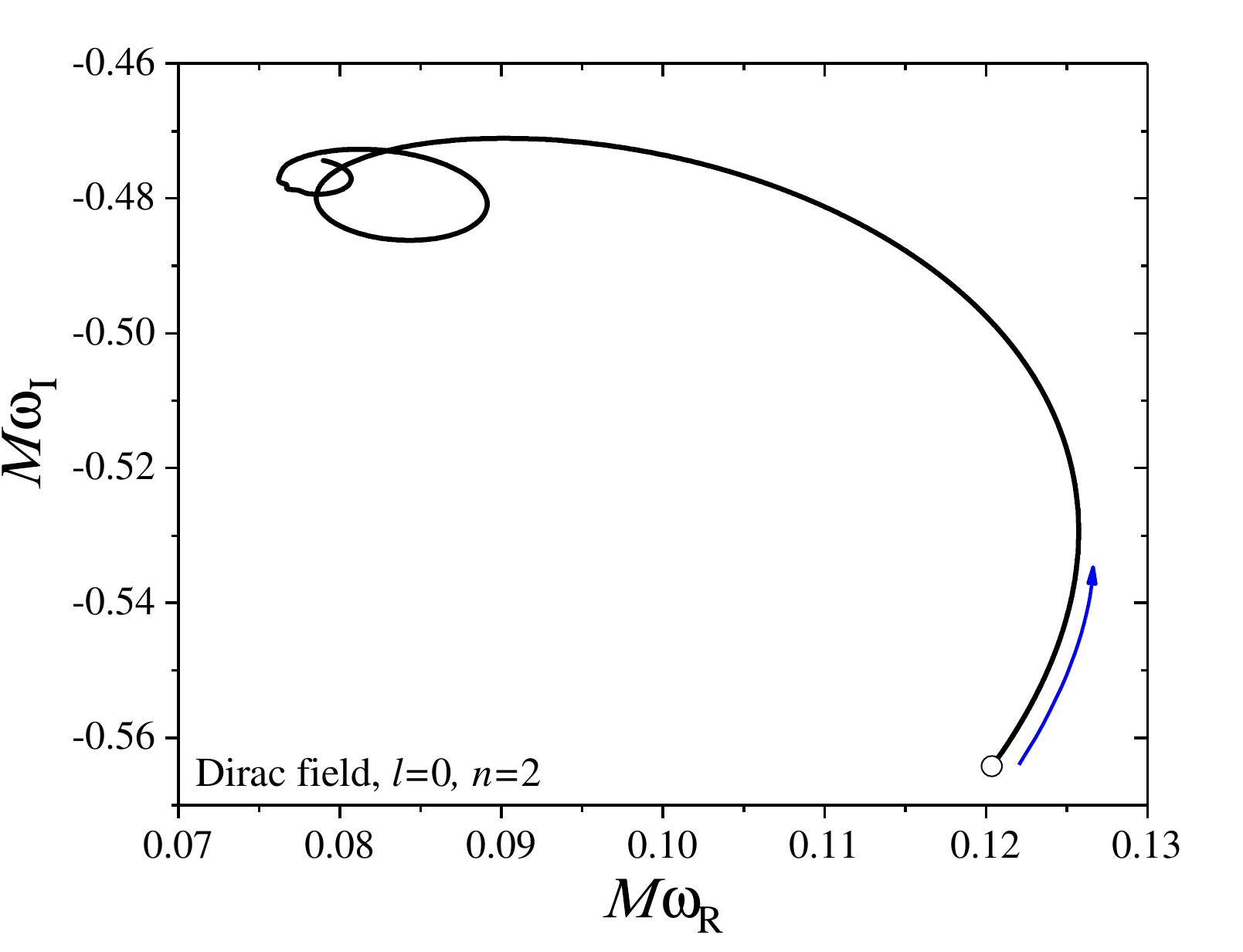}
       \caption{
The detailed behavior of the first  ($n=1$)  overtone for the scalar (top panel) and Dirac (bottom panel) fields with  $l=0$.}
    \label{fig:Sf_Df_l=0}
\end{figure}

\begin{figure}
    \centering
    \includegraphics[width=.49\textwidth]{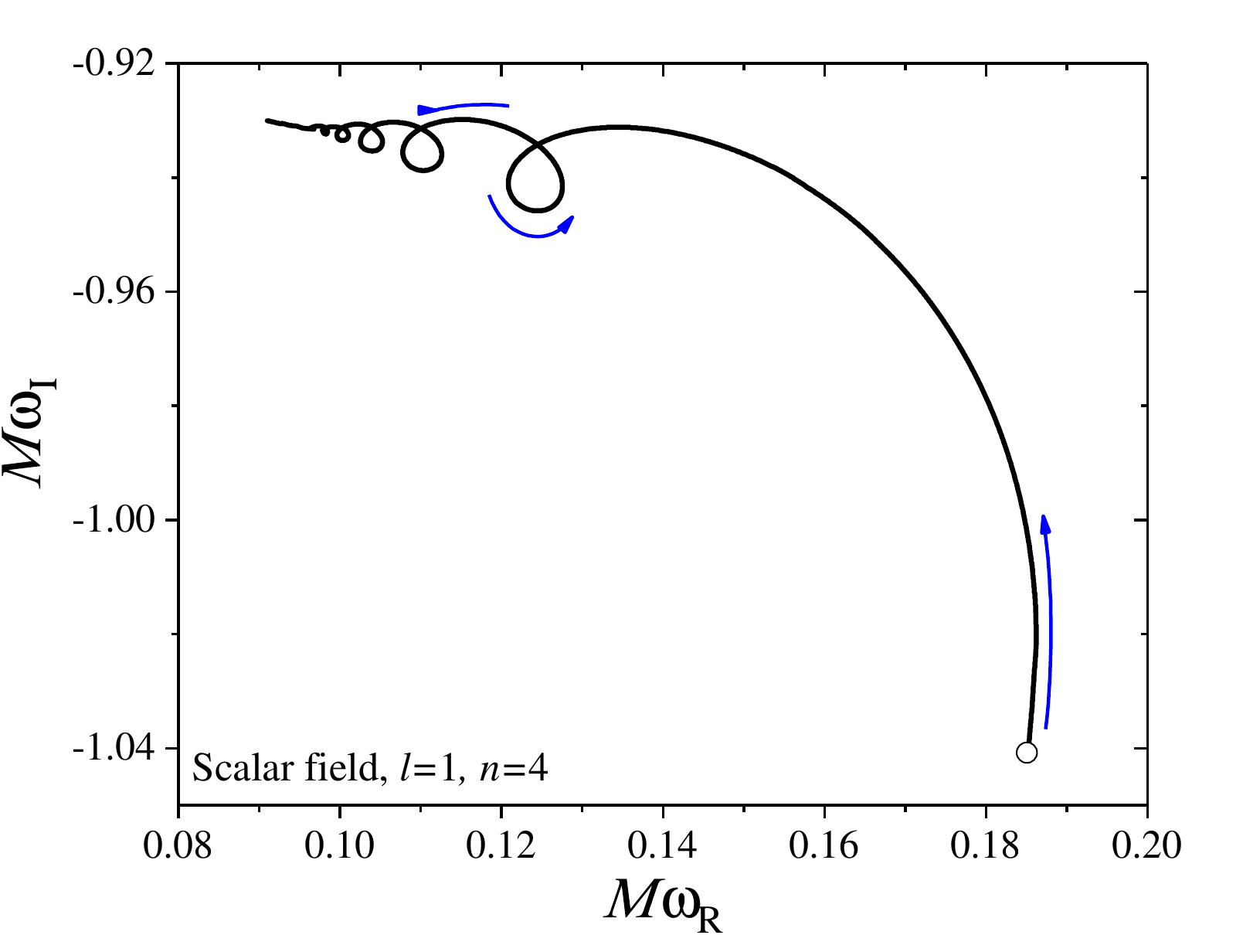}
    \includegraphics[width=.49\textwidth]{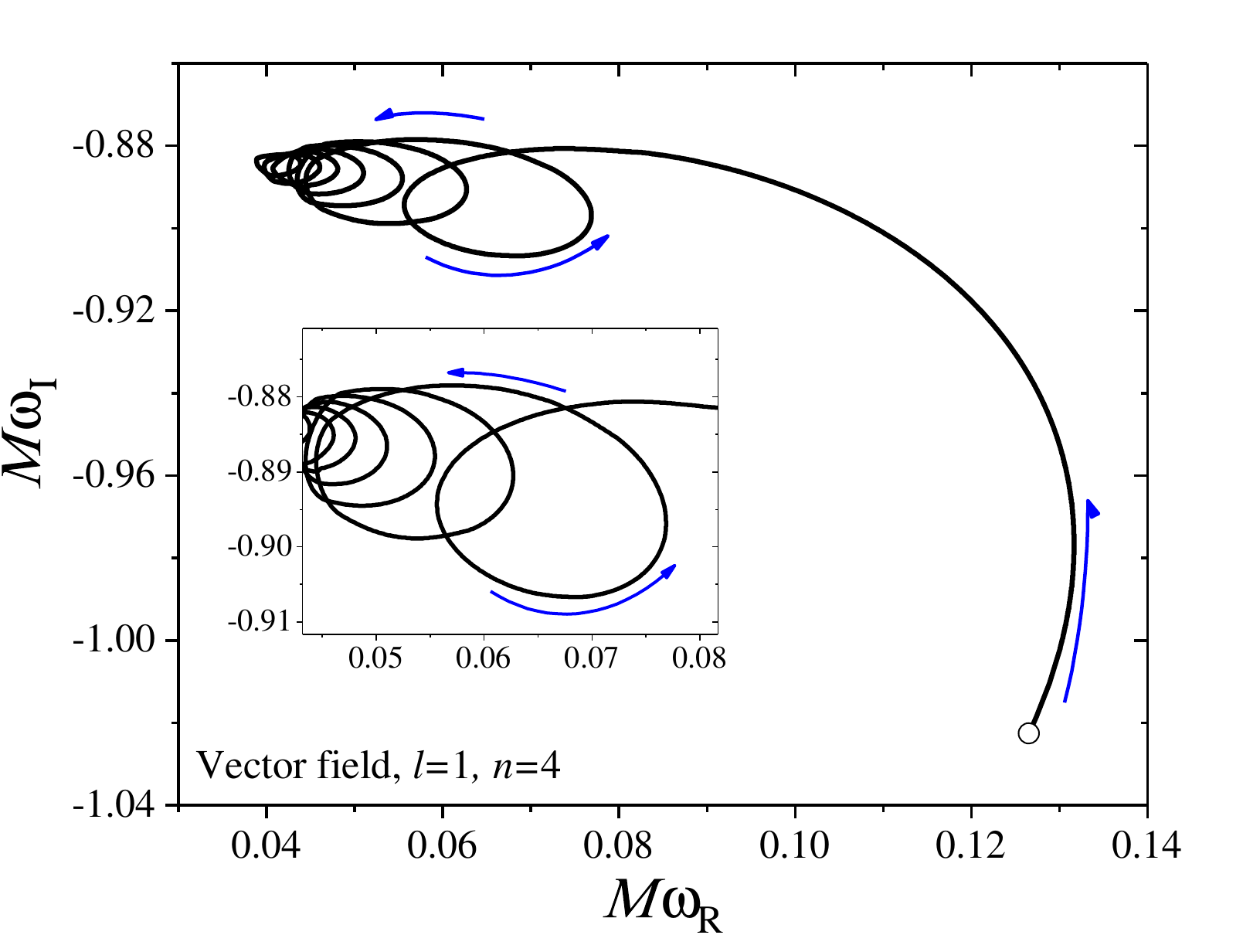}
    \caption{The detailed behavior of  $n=4$ overtone for the scalar (top panel) and vector (bottom panel) fields with  $l=1$.}
\label{fig:overton_example}
\end{figure}

The typical behaviour of specific overtones are shown in Fig. \ref{fig:Sf_Df_l=0} for the scalar and Dirac fields for $l=0$ and in Fig. \ref{fig:overton_example} for the scalar and vector fields with $l=1$.
This spiral-like behaviour  is more pronounced in the case of the vector field, appearing from $
n=3$ for $l=1$, while in the case of the Dirac field it is less pronounced, with spirals appearing from $n=2$ for $l=0$ and  from $n=5$ for $l=1$. For the scalar field, such behavior appears from $n=1$ for $l=0$ and from $n=4$ for $l=1$. 
This trend remains for higher values of $l$. Increasing $l$ leads to increasing  $n$ where such spirals can be observed, the number of spiral turns increases with increasing $n$.

\section{Black hole shadows and eikonal approximation}
\label{sec:Eikonal}
In this section, we investigate the properties of null geodesics and the eikonal approximation for quasinormal modes. We begin with the Lagrangian for geodesics, given by
\eq{\mathcal {L} \equiv \dfrac{1}{2}g_{\mu \nu }\dot{x}^{\mu }\dot{x}^{\nu },
}
where $\dot{x}^{\mu }=dx^{\mu }/d\tau$ and $\tau$ is an affine parameter along the geodesics. 
Without loss of generality, we can restrict our consideration to geodesic motion in the equatorial plane. The metric (\ref{eq:metric})  has two Killing vectors $\partial _t$ and $\partial_\phi$, leading to two integrals of motion
\eq{
f(r)\dot{t}=E,\quad r^2\dot{\phi}=L,
}
which correspond to the energy $E$ and angular momentum $L$ of the particle, respectively. Together with the additional integral of motion for massless particles,  $g_{\mu \nu }\dot{x}^{\mu }\dot{x}^{\nu }=0$,  we obtain 
\eq{\label{eq:Ugeod}
(\dot{r})^2=E^2-U_{\rm eff}(r),\quad U_{\rm eff}(r)=f(r)\frac{L^2}{r^2},}
where $U_{\rm eff}(r)$ denotes the effective radial potential of the geodesic motion. We are interested in photon circular orbits, which satisfy the conditions
\eq{
U_{\rm eff}(r)=E^2,\quad U_{\rm eff}'(r)=0,
}
that yield
\eq{\label{eq:photonGeod}
rf'(r)-2f(r)=0.}
Substituting the metric function (\ref{eq:f(r)}) into (\ref{eq:photonGeod})  we get the equation defining a radius $r_{\rm ph}$ of the photon sphere 
\eq{r^3-3Mr^2+6M\xi=0.}
For $\xi\in[0,\xi_{\rm cr}]$, this equation has 3 roots. Two of these roots are less than the outer horizon radius, while the larger root corresponds to the photon sphere and is given by
\eq{\label{eq:rad_ph_sphere}
r_{\rm ph}/M=1+2 \sin \left[\frac{1}{3} \arccos\left( \frac{3\xi}{M^2}
   -1\right)+\frac{\pi}{6} \right],
}
The observed radius of the photon sphere, due to gravitational lensing, appears to be a circle with a radius of \cite{PhysRevLett.125.141104}
\eq{\label{eq:shadow_radii}
r_{\rm sh}=\frac{r_{\rm ph}}{\sqrt{f(r_{\rm ph})}}.
}
One can see that $r_{\rm sh}(\xi)$ is a monotonically decreasing function.

Using (\ref{eq:shadow_radii}), we obtain the following interval
\eq{
4.869 \lesssim r_{\rm sh}/M\lesssim 5.196,~~\xi\in[0,\xi_{\rm cr}],
}
which aligns with recent EHT observations for  $M87^*$ $r_{\rm sh}/M\in(4.31,6.08)$  \cite{PhysRevLett.125.141104} and for Sgt $A^*$, where  $r_{\rm sh}/M\in(4.55, 5.22)$ for $1\sigma$ and $r_{\rm sh}/M\in (4.21,5.56)$ for $2\sigma$ \cite{Vagnozzi_2023}.

There is a strong correlation between null geodesics and QNMs of static black holes \cite{PhysRevD.79.064016}, as well as for rotating one \cite{Jusufi_2020,Yang_2021, 2024arXiv240407589P}.
In the eikonal limit  ($l\to\infty$), the effective potential for all spins has the following form
\eq{
V_{\rm eff}(r)\simeq f(r)\frac{l^2}{r^2}+O\left(\frac{1}{l}\right),
}
which is coincides with the effective potential for the geodesic motion (\ref{eq:Ugeod}). 

The QNM frequencies can be written as
\eq{\label{eq:W_eik}
\omega=l\Omega-i\left(n+\frac{1}{2}\right)\lambda,
}
where $\Omega$ and $\lambda$ are the angular velocity and Lyapunov exponent of photons on the photon orbit, respectively. 
They have the following form  \cite{PhysRevD.79.064016}
\eq{\label{eq:Omega_lambda}
\Omega=\frac{\sqrt{f(r)}}{r},~\lambda=r\sqrt{-\frac{f(r)}{2}\frac{d^2}{dr^2}\left(\frac{f(r)}{r^2}\right),}
}
evaluated at $r=r_{\rm ph}.$

For small $\xi$, the compact form of (\ref{eq:W_eik}) can be obtained as
\eq{
\Omega=\frac{1}{3 \sqrt{3}M}+\frac{\xi }{27
   \sqrt{3}M^3}+\frac{25 \xi ^2}{1458 \sqrt{3}M^5}+O(\xi^3),
   }
\eq{
\lambda=\frac{1}{3 \sqrt{3}M}-\frac{2 \xi }{27
   \sqrt{3}M^3}-\frac{46 \xi ^2}{729 \sqrt{3}M^5}+O(\xi^3),
}
when we observe that $\omega_{\rm R}$ and $\omega_{\rm I}$  are larger and smaller, respectively, than in the Schwarzschild black hole case.

\section{Scattering problem and grey-body factors}
\label{sec:scattering}
The black hole grey-body factors modify the spectrum of Hawking radiation observed at spatial infinity by a distant observer, determining the proportion of initial quantum radiation reflected back to the event horizon by the potential barrier. 

The scattering boundary conditions have the following form
\eq{\label{BC}
 \Psi\sim e^{-i\omega r_*} + R e^{i\omega r_*},\,r\to{r_{\rm h}},\,\,\Psi\sim T e^{-i\omega r_*},\,r\to{\infty}
}
where $\omega$ is real, and $R$ and $T$ correspond to the reflection and transmission coefficients, respectively. They satisfy
\begin{equation}\label{1}
\left|T\right|^2 + \left|R\right|^2 = 1.
\end{equation}

The black hole grey-body factor $A_l$ is defined as  the transmission coefficient
\begin{equation}
A_l\equiv \left|T\right|^2=1-\left|R\right|^2.
\end{equation}
The  transmission for each multipole  $l$ can be calculated using the WKB approach 
\begin{equation}\label{moderate-omega-wkb_1}
|R|^2 = \frac{1}{1 + e^{- 2 i \pi K}},\quad |T|^2=\frac{1}{1 + e^{ 2 i \pi K}}.
\end{equation}
The phase factor $K$ is determined using  equation  \cite{Konoplya_2019}
\eq{ K=
\frac{i\left(\omega^2 - V^{(i)}_{\rm eff}(r_m)\right)}{\sqrt{-2\partial^2_{(r^*)^2} V^{(i)}_{\rm eff}(r_m)}} + \sum_{i=2}^{6} \Lambda_{i}(K) ,
}
where $r_m$ is the point of the maximum of the effective potential $V^{(i)}_{\rm eff}$ and $\Lambda_i(K)$ corresponds to higher-order WKB corrections \cite{Konoplya_2019,Konoplya_2003} at $r=r_m$. 

\begin{figure*}
    \centering
    \includegraphics[width=1\textwidth]{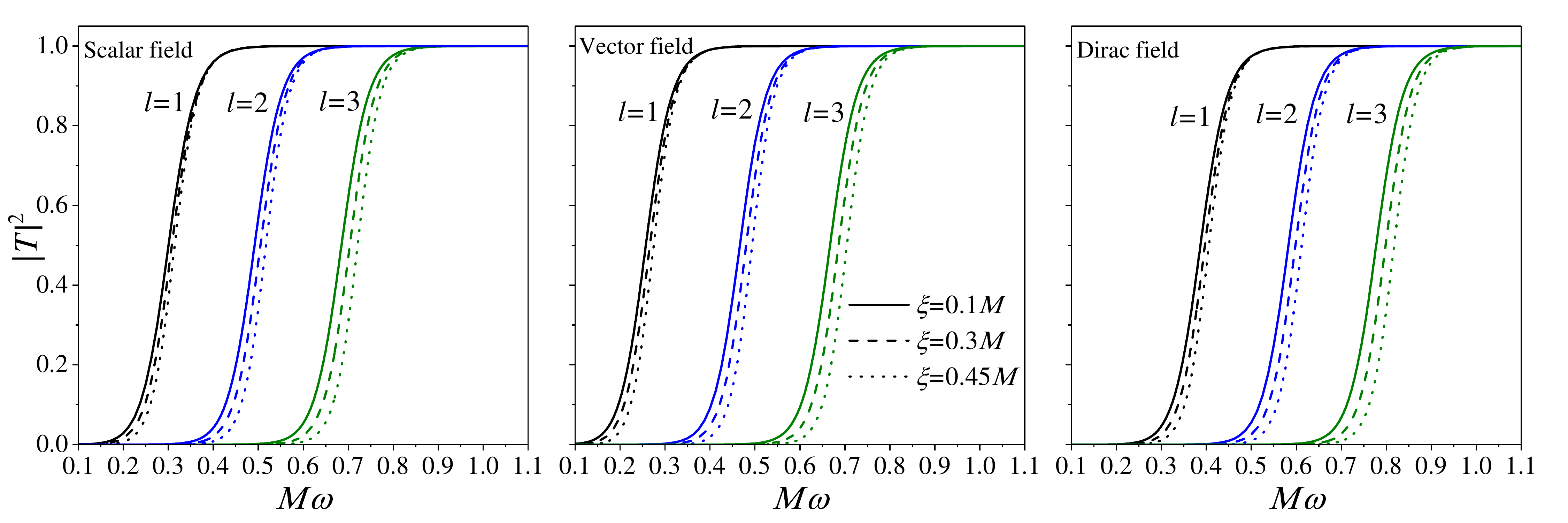}
\caption{Typical examples of the gray-body factors for the scalar, vector and Dirac fields, respectively }
\label{fig:GraybodyFactors}
\end{figure*}

Typical examples of grey-body factors as a function of  $\omega$ are shown in Fig. \ref{fig:GraybodyFactors}.  One can observe that grey-body factors decrease with increasing $\xi$, which means that  a larger fraction of particles is reflected by the effective potential.

Following \cite{PhysRevA.59.427,2008AnPhy.323.2779B}, we have the lower bounds on the grey-body factors. They are given by
\eq{
A_l \geq \sech^2\left[ \int\limits_{-\infty}^{\infty}  \Xi(r(r^*))  d r^* \right],
}
where $\Xi(r)$ is the following function
\begin{equation}
\Xi(r) = {\sqrt{ (h')^2 + [\omega^2-V^{(i)}_{\rm eff}(r)- h^2]^2}\over2 h }.
\label{E:b0}
\end{equation}
and $h(r^*)>0$ satisfies $h(-\infty)=h(+\infty)=\omega$. 

Taking  $h=\omega$, we have
\eq{\label{eq:GB_LB}
A_l \geq \sech^2\left[ {1\over2\omega} \int_{-\infty}^{\infty} V^{(i)}_{\rm eff}(r(r^*))  d r^* \right],
}
We can obtain the analytical expression of the lower bounds of the grey-body factors. After substituting the exact form of $V^{(i)}_{\rm eff}$, $i=(s,v,d)$ in (\ref{eq:GB_LB}), we obtain 
\eq{
A_l^{(s)}(\xi)\geq \sech^2\left[\frac{2l(l+1)+1}{4\omega r_h(\xi)}+\frac{z(\xi)}{2\omega}\right],
}
where $z(\xi)$ has the following form
\begin{equation}
\begin{split}
&z(\xi)=\frac{\sqrt{3}M \pi }{a^2}+\frac{3}{2r_h}-\frac{6M}{r_h^2}
   {}_2F_1\left(\frac{2}{3},1,\frac{5}{3},-\frac{6M\xi}{r_h^3}\right)\\
   &-\frac{2 M\sqrt{3}}{a^2} \arctan\left(\frac{2 r_h-a}{\sqrt{3}
   a}\right)-\frac{M}{a^2}\log \left(\frac{(a+r_h)^2}{(a-r_h)^2+a r_h}\right),
\end{split}
\end{equation}
where $a=(6M\xi)^{1/3}$ and $r_h=r_h(\xi)$. For small $\xi$, we have
\eq{
z(\xi)=-\frac{\sqrt{3} \pi  a}{128 M^2}-\frac{3 \sqrt{3} \pi  a^2}{2048 M^3}+O\left(a^3\right),
}
In the case of vector and Dirac fields, the corresponding expressions have a more compact form
\eq{
A_l^{(v)}(\xi)\geq \sech^2\left[\frac{l(l+1)}{2\omega r_h(\xi)}\right],
}

\eq{
A_l^{(d)}(\xi)\geq \sech^2\left[\frac{(l+1)^2}{2\omega r_h(\xi)}\right], 
}
One can see that $r_h(\xi)$ and $z(\xi)$  are monotonically decreasing functions with $r_h(0)=2M$, $r_h(\xi_{\rm cr})=r_h^{\rm ext}<2M$ and $z(0)=0$, $z(\xi_{\rm cr})<0$. Consequently, $A_l^{(s,v,d)}(\xi)\leq  A_l^{(s,v,d)}(0)$ and the lower bounds for  grey-body factors are always smaller compared to those for the Schwarzschild black hole. 
\section{Conclusions}
\label{sec:conclusions}

In this paper, we provide a detailed consideration of the QNM spectra of massless scalar, vector, and Dirac fields for the regular black hole spacetime described by  (\ref{eq:f(r)})  which corresponds to the static exterior of a collapsing dust ball within the framework of asymptotically safe gravity \cite{Bonanno_etal2024PhRvL}. We obtain accurate values for the fundamental mode and the first few overtones for each test field.

We find that the fundamental mode is weakly sensitive to the scale parameter $\xi$,  but for the overtones we have  the peculiar spiral-like behavior in the QNM trajectories in the complex $\omega$ plane  as a function of $\xi$.  

Additionally, we study black hole shadow and demonstrate that shadow radius  decreases with increasing $\xi$, and  for $\xi\in(0,\xi_{\rm cr}]$ it aligns with current EHT results. In the context of the eikonal approximation related to black hole shadows, we find that the real and imaginary parts of $\omega$ are always larger and smaller, respectively, than in the Schwarzschild black hole case. 

Furthermore, we study the grey-body factors and derive explicit lower bound estimations. We find that they decrease with increasing $\xi$ and they are always smaller compared to those in the case of the Schwarzschild black hole.

\FloatBarrier

\vskip3mm \textit{Acknowledgements.}\\
O.S. is grateful to Valery Zhdanov for fruitful discussions. 
\FloatBarrier
\appendix
\section{Accurate values of QNM frequencies}
\begin{table*}[ht]
\caption{The accurate values of QNM frequencies for the fundamental mode and the first three overtones for the scalar field perturbations.}
\label{tab:scalar}
\begin{tabular}{l|c|c|c|c|c}
   \hline
 \multicolumn{6}{c}{Scalar field}\\
        \hline
$\xi$ & $n$ & $l=0$    & $l=1$    & $l=2$&$l=3$\\ 
\hline
  $ 0$ & \makecell{ $ 0 $ \\ $ 1 $\\ $ 2$\\ $3$ }     &     \makecell{ $ 0.1104549-0.1048957 i$ \\
 $0.0861169-0.3480524 i$ \\
 $0.0757396-0.6010787 i$ \\
 $0.0704483-0.8536819 i$ }& \makecell{ $ 0.2929361-0.0976600 i$ \\
 $0.2644487-0.3062574 i$ \\
 $0.2295393-0.5401334 i$ \\
 $0.2032584-0.7882978 i$  } & \makecell{ $ 0.4836439-0.0967588 i$ \\
 $0.4638506-0.2956039 i$ \\
 $0.4305441-0.5085584 i$ \\
 $0.3938631-0.7380966 i$ } & \makecell{ $0.6753662-0.0964996 i$ \\
 $0.6606715-0.2922848 i$ \\
 $0.6336258-0.4960082 i$ \\
 $0.5987733-0.7112212 i$} \\
   \hline
  $ 0.1$  & \makecell{ $ 0 $ \\ $ 1 $\\ $ 2$\\ $3$ }   &     \makecell{ $ 0.1127147 - 0.1017380i$ \\ $ 0.0879157 - 0.3352088i$\\
  $0.0733768 - 0.5790211 i$\\
  $0.0615 - 0.8236 i$\\}& \makecell{ $ 0.2966692 - 
0.0950633i$ \\ $0.2701353 - 
0.2967061i$\\
 $0.2358748 - 0.5206370i$\\
 $0.2073836 - 0.7580274i$} & \makecell{ $ 0.4894369-0.0943001 i$ \\
 $0.4711135-0.2875314 i$ \\
 $0.4398335-0.4930068 i$ \\
 $0.4041963-0.7130744 i$ }&\makecell{$0.6833396-0.0940930 i$ \\
 $0.6697399-0.2847045 i$ \\
 $0.6445627-0.4822058 i$ \\
 $0.6116627-0.6896842 i$ } \\
   \hline
  $ 0.2$   &\makecell{ $ 0 $ \\ $ 1 $\\ $ 2$\\ $3$ }  &     \makecell{ $ 0.1146069 - 0.0977301i$ \\
  $0.0863938 - 0.3192448i$ \\$0.05862 - 0.55266 i$\\
  $-$}& \makecell{ $ 0.3005978 - 0.0919319i$ \\ $0.2752068 - 0.2852384i $\\
  $0.2391871 - 0.4972741i$\\
  $0.2033367 - 0.7217116i$} & \makecell{ $  0.4957590-0.0913311 $ \\
 $0.4785043-0.2777928i$ \\
 $0.4482205-0.4743052 i$ \\
 $0.4116051-0.6830412 i$ \\} & \makecell{ $0.6921221-0.0911781 i$ \\
 $0.6793563-0.2755264 i$ \\
 $0.6554330-0.4655213 i$ \\
 $0.6233622-0.6637171 i$}  \\
   \hline
  $ 0.3$   &\makecell{ $ 0 $ \\ $ 1 $\\ $ 2$\\ $3$ }  &     \makecell{ $0.1154763 - 0.0925994i$ \\ $ 0.0737228 - 0.3002834i$\\
  $0.0420 - 0.5783 i$\\$0.036 - 0.806i$ }& \makecell{ $ 0.3046570 - 
 0.0880529i$ \\ $ 0.2785571- 
0.2712995i$\\
 $0.2349018 - 0.4700198i$\\
 $0.1759422 - 0.6868101i$} & \makecell{ $0.5026982 - 0.0876222i $ \\ $ 0.4856091 - 0.2657174i $ \\ $0.4540345 -0.4514723i$\\$0.4118267 - 
 0.6472222i$ } & \makecell{ $0.7019046-0.0875206 i$ \\
 $0.6893740-0.2640535 i$ \\
 $0.6653242-0.4448326 i$ \\
 $0.6315770-0.6319037 i$} \\
 \hline
  $ 0.4$  &\makecell{ $ 0 $ \\ $ 1 $\\ $ 2$\\ $3$ }   &     \makecell{ $ 0.1132129 - 0.0870447i$ \\ $ 0.0672842 - 0.3059916i$\\
  $0.038 - 0.549 i$\\$-$}& \makecell{ $ 0.308589 - 
0.0831396i$ \\
$0.2776680 - 0.2555897i$\\
 $0.2192663 - 0.4494627i$\\
 $0.1529278 - 0.6786560i$} & \makecell{ $ 0.5103010
 - 0.0827738i $ \\ $ 0.4911978 - 0.2505045i $ \\ $0.4533703 - 
 0.4250037i$ \\ $0.3979061 - 0.6120243i$ } & \makecell{ $ 0.7129261-0.0826928 i$ \\
 $0.6991385-0.2491847 i$ \\
 $0.6717046-0.4190342 i$ \\
 $0.6308843-0.5946843 i$ }  \\
 \hline
  $ 0.45$  &\makecell{ $ 0 $ \\ $ 1 $\\ $ 2$\\ $3$ }   &     \makecell{ $ 0.1114437 - 0.0863403i$ \\ $0.0616756 - 0.3055581i$\\$0.034 - 0.555i$\\$-$ }& \makecell{ $ 0.3103227 - 0.0802535i$ \\
$0.2753474-0.2483725i$\\$0.2115478 - 0.4428928i$\\$0.1436060 - 0.6754821i$} & \makecell{ $ 0.5143087- 0.0797253i $ \\ $ 0.4926241 - 0.2417374i $\\ $0.4495238 - 0.4121991i$\\ $0.3873515-0.5988040i$ } & \makecell{ $0.7189708-0.0796013 i$ \\
 $0.7033408-0.2400759 i$ \\
 $0.6720733-0.4045739 i$ \\
 $0.6254514-0.5764418 i$ }  \\
 \hline
$ \xi_{\rm cr}$ &\makecell{ $ 0 $ \\ $ 1 $\\ $ 2$\\ $3$ }    &     \makecell{ $ 0.1113158 -0.0862468i $ \\
$0.0614446 - 0.3053450 i$\\$-$\\$-$ }& \makecell{ $ 0.3105288-0.0798624i$ \\
$0.2749861-0.2474982i$\\
$0.210590 - 0.4421533i$\\$0.1424936 - 0.6750934 i$ } & \makecell{ $ 0.5148323-0.0792934i $ \\ $ 0.4927130 - 0.2405676i $\\ $
0.4488801-0.4106422i$\\$
0.3859468 - 0.5973218i$ } & \makecell{ $ 0.7197787-0.0791577 i$ \\
 $0.7038180-0.2388113 i $\\
 $0.6719382-0.4026820 i$ \\
 $0.6245253-0.5742208 i $ } \\
   \hline

\end{tabular}
\end{table*}

\begin{table*}[]
\label{tab:Vector}
\caption{The accurate values of QNM frequencies for the fundamental mode and the first three overtones for the vector field perturbations.}
\begin{tabular}{l|c|c|c|c|c}
   \hline
 \multicolumn{6}{c}{Vector field}\\
        \hline
$\xi$ & $n$    & $l=1$    & $l=2$&$l=3$& $l=4$\\ 
\hline
  $ 0$  & \makecell{ $ 0 $ \\ $ 1 $\\ $ 2$\\ $3$ }   &     \makecell{ $  0.2482633-0.0924877 i$ \\
$ 0.2145154-0.2936676 i$ \\
 $0.1747736-0.5251876 i$ \\
 $0.1461769-0.7719095 i $ }& \makecell{ $ 0.4575955-0.0950044 i$ \\
 $0.4365424-0.2907101 i$ \\
 $0.4011867-0.5015873 i $\\
 $0.3625950-0.7301985 i$ } & \makecell{ $ 0.6568987-0.0956162 i$ \\
 $0.6417374-0.2897284 i$ \\
$ 0.6138320-0.4920663 i$ \\
$ 0.5779185-0.7063308 i$ }&\makecell{$ 
 0.8530952-0.0958599 i$ \\
 $0.8412671-0.2893147 i$ \\
 $0.8187284-0.4878383 i $\\
 $0.7877483-0.6942424 i$}  \\
   \hline
  $ 0.1$ & \makecell{ $ 0 $ \\ $ 1 $\\ $ 2$\\ $3$ }    &     \makecell{ $  0.2529550-0.0901305 i$ \\
 $0.2221777-0.2844042 i$ \\
 $0.1842388-0.5052108 i$ \\
 $0.1544984-0.7402078 i$\\}& \makecell{
 $0.4638067-0.0925696 i $\\
 $0.4445106-0.2826667 i$ \\
 $0.4117074-0.4859211 i$ \\
 $0.3747812-0.7047553 i$} & \makecell{ $ 0.6651354-0.0932132 i$ \\
 $0.6511790-0.2821492 i$ \\
 $0.6253580-0.4782281 i $\\
$ 0.5916936-0.6846612 i$ }  & \makecell{ $ 0.8634757-0.0934777 i $\\
 $0.8525631-0.2819456 i$ \\
$ 0.8317104-0.4748207 i$ \\
$ 0.8028601-0.6745298 i$ }  \\
   \hline
  $ 0.2$  & \makecell{ $ 0 $ \\ $ 1 $\\ $ 2$\\ $3$ }   &     \makecell{  $0.2580323-0.0871284 i$ \\
 $0.2296843-0.2728136 i$ \\
 $0.1916661-0.4805530 i$ \\
 $0.1567062-0.7011726 i$}& \makecell{ $ 0.4706321-0.0895915 i$ \\
 $0.4527304-0.2728391 i$ \\
 $0.4215152-0.4668556 i$ \\
 $0.3843276-0.6738614 i$} & \makecell{ $  0.6742384-0.0902859 i $\\
$ 0.6612387-0.2729190 i$ \\
$ 0.6369107-0.4614025 i $\\
$ 0.6044028-0.6583838 i $ } &\makecell{ $0.8749729-0.0905776 i$ \\
 $0.8647880-0.2729758 i$ \\
 $0.8451990-0.4589877 i$ \\
 $0.8177396-0.6505947 i$} \\
   \hline
  $ 0.3$  & \makecell{ $ 0 $ \\ $ 1 $\\ $ 2$\\ $3$ }   &     \makecell{ $ 0.2634890-0.0831610 i$ \\
 $0.2359680-0.2578023 i$ \\
 $0.1925527-0.4492398 i$ \\
 $0.1377966-0.6534130 i$ }& \makecell{ $ 0.4781936-0.0858092 i$ \\
 $0.4608113-0.2604402 i$ \\
 $0.4289299-0.4431336 i$ \\
 $0.3868825-0.6361283 i$ } & \makecell{ $  0.6844228-0.0865847 i$ \\
 $0.6717905-0.2612895 i$ \\
 $0.6475881-0.4403633 i$\\
 $0.6137403-0.6258935 i$ } &\makecell{  $0.8878862-0.0869144 i$ \\
 $0.8779829-0.2616703 i$ \\
 $0.8586760-0.4391236 i$ \\
 $0.8309201-0.6207804 i$ } \\
 \hline
  $ 0.4$   & \makecell{ $ 0 $ \\ $ 1 $\\ $ 2$\\ $3$ }  &     \makecell{ $ 0.2690795-0.0776770 i$ \\
 $0.2374610-0.2389323 i $\\
 $0.1762254-0.4210579 i $\\
 $0.1060480-0.6417165 i$}& \makecell{ $ 0.4865834-0.0807453 i$ \\
 $0.4674302-0.2444092 i $\\
 $0.4295015-0.4147645 i$ \\
 $0.3737084-0.5974858 i $} & \makecell{ $ 0.6959663-0.0816441 i$ \\
 $0.6821753-0.2460390 i $\\
 $0.6547398-0.4137863 i $\\
 $0.6139137-0.5873080 i $    } &\makecell{ $0.9026457-0.0820267 i$ \\
$ 0.8918706-0.2467472 i$ \\
$ 0.8703912-0.4134791 i$ \\
 $0.8383335-0.5836122 i$} \\
 \hline
  $ 0.45$  & \makecell{ $ 0 $ \\ $ 1 $\\ $ 2$\\ $3$ }   &     \makecell{ $ 0.2716327-0.0741932 i$ \\
 $0.2351787-0.2298792 i$ \\
 $0.1671128-0.4127434 i$ \\
 $0.0943476-0.6367767 i $ }& \makecell{ $ 0.4910531-0.0774822 i$ \\
 $0.4692235-0.2349643 i $\\
 $0.4256956-0.4008310 i $\\
 $0.3626391-0.5829953 i $} & \makecell{ $  0.7023330-0.0784421 i$ \\
 $0.6866741-0.2365882 i$ \\
$ 0.6553231-0.3987356 i$ \\
 $0.6085163-0.5682475 i $ } & \makecell{ $0.9108950-0.0788507 i$ \\
$ 0.8986799-0.2373078 i$ \\
$ 0.8742202-0.3981100 i$ \\
$ 0.8375353-0.5630788 i$} \\
 \hline
$ \xi_{\rm cr}$  & \makecell{ $ 0 $ \\ $ 1 $\\ $ 2$\\ $3$ }   &     \makecell{ $ 0.2719364-0.0737089 i$ \\
 $0.2347808-0.2287955 i$ \\
 $0.1659914-0.4118425 i$ \\
$ 0.0930899-0.6362850 i$ }& \makecell{ $ 0.4916390-0.0770148 i$ \\
$ 0.4693433-0.2336966 i$ \\
$ 0.4250286-0.3991411 i$ \\
$ 0.3611578-0.5813853 i $ } & \makecell{ $ 0.7031859-0.0779799 i$ \\
 $0.6871848-0.2352702 i$ \\
 $0.6551999-0.3967625 i$ \\
$ 0.6075738-0.5659294 i $ } &\makecell{ $0.9120094-0.0783908 i$ \\
$ 0.8995251-0.2359695 i$ \\
$ 0.8745500-0.3960160 i$ \\
$ 0.8371519-0.5604263 i$ } \\
   \hline

\end{tabular}
\end{table*}
\appendix
\begin{table*}[]
\caption{The accurate values of QNM frequencies for the fundamental mode and the first three overtones for the Dirac field perturbations.}
\label{tab:dirac}
\begin{tabular}{l|c|c|c|c|c}
   \hline
 \multicolumn{6}{c}{Dirac field}\\
        \hline
$\xi$ & $n$ & $l=0$    & $l=1$    & $l=2$& $l=3$\\ 
\hline
  $ 0$  & \makecell{ $ 0 $ \\ $ 1 $\\ $ 2$\\ $3$ }  &     \makecell{ $ 0.1829629-0.0969824 i$ \\
 $0.1478222-0.3169284 i$ \\
 $0.1203701-0.5642265 i$ \\
 $0.1042576-0.8169780 i$ }& \makecell{ $ 0.3800801 - 0.0963821i $ \\ $ 0.3558964 - 0.2974182i$\\
  $0.3193482 - 0.5182988i$\\
  $0.2850289 - 0.7571955i$} & \makecell{ $0.5740940-0.0963048 i$ \\
 $0.5570155-0.2927154 i$ \\
 $0.5266070-0.4996948 i$ \\
 $0.4897451-0.7208783 i$ } & \makecell{$0.7673546-0.0962699 i$ \\
 $0.7543005-0.2909679 i$ \\
 $0.7297698-0.4919098 i$ \\
 $0.6969131-0.7022932 i$} \\
   \hline
  $ 0.1$  & \makecell{ $ 0 $ \\ $ 1 $\\ $ 2$\\ $3$ }  &     \makecell{ $ 0.1854229 - 0.0940140i$ \\ $ 0.1528740 - 0.3046905i$\\
  $0.1252016 - 0.5401493i$\\
  $0.1067 - 0.7810i$\\}& \makecell{ $ 0.3845602 - 
 0.0938853i$ \\ $0.3621899 - 0.2888616i$\\
 $0.3275457 - 0.5011678i$\\
 $0.2932295 - 0.7297184i$} & \makecell{ $ 0.5808381 - 
 0.0938816i $ \\ $ 0.5650414 - 
0.2849551i$\\ $0.5366717 - 0.4852048i$ \\ $0.5015528 - 0.6978452i$ } & \makecell{$0.7763373 - 
0.0938821i$\\ $0.7642566 - 
0.2835269i$\\ $0.7414621 - 
0.4785965i$\\ $0.7106380 - 
0.6818565i$} \\
   \hline
  $ 0.2$   & \makecell{ $ 0 $ \\ $ 1 $\\ $ 2$\\ $3$ }  &     \makecell{ $ 0.1879348 - 0.0904495i$ \\
  $0.1564867 - 0.2899489i$ \\$0.1239379 - 0.5107052i$\\
  $0.0933 - 0.7359i$}& \makecell{ $ 0.3894888 - 0.0908362i$ \\ $0.3684325 - 0.2784193i $\\
  $0.3342687 - 0.4803017i$\\
  $0.2969723 - 0.6961613i$} & \makecell{ $ 0.5882679 - 
 0.0909399i$ \\ $ 0.5734486 - 
0.2755380i $\\ $ 0.5463680 - 
0.4676571i $\\ $ 0.5115672 - 0.6700156i $ }  &\makecell{$0.786270-0.090979i$\\$0.7749458 - 
0.2744840i$\\$0.7533903 - 
0.4624345i$\\$0.7237112 - 
0.6570981i$\\}\\
   \hline
  $ 0.3$   & \makecell{ $ 0 $ \\ $ 1 $\\ $ 2$\\ $3$ } &     \makecell{ $0.1903278 - 0.0860101i$ \\ $ 0.1564381 - 0.2718541i$\\$0.103908 - 0.475866i$\\$-$ }& \makecell{ $ 0.3948574 - 
 0.0870123i $ \\ $ 0.3739159 - 0.2654483i$\\
 $0.3368955 - 0.4548696i$\\
 $0.2891317 - 0.6566533i$} & \makecell{ $ 0.5965315 - 
0.0872385i $ \\ $ 0.5819776 - 
0.2637447i $ \\$ 0.5544731- 
0.4459028i $ \\$ 0.5167316 - 0.6360134i$  } & \makecell{ $0.7973937 - 
0.0873226i$ \\ $ 0.7863156 - 
0.2631210i $\\ $ 0.7648531 - 
0.4422451i $\\ $ 0.7343024 - 
0.6264460i $ } \\
 \hline
  $ 0.4$   & \makecell{ $ 0 $ \\ $ 1 $\\ $ 2$\\ $3$ } &     \makecell{ $ 0.1919608 - 0.08037408i$ \\ $ 0.1465229 - 0.2548618i$\\
  $0.07673-0.4781i$\\$-$}& \makecell{ $ 0.4005913 - 
 0.0819991i $ \\
 $ 0.3767973 - 0.2493739i$\\
 $0.3298911-0.4275506i$\\
 $0.2636033 - 0.6271177i$} & \makecell{ $ 0.6057860 - 
0.0823403i $ \\ $ 0.5896983 - 
0.2485171i $\\ $ 0.5577336 - 0.4192578i $\\ $ 0.5102957 - 0.5980576i $ } &\makecell{$0.8100370 - 
0.0824657i$\\$0.7979026 - 
0.2482363i$\\$0.7737293 - 
0.4165427i$\\ $0.7376859 - 0.5891690i$
}  \\
 \hline
  $ 0.45$   & \makecell{ $ 0 $ \\ $ 1 $\\ $ 2$\\ $3$ } &     \makecell{ $ 0.1920794 - 0.0773254i$ \\ $ -$\\$-$\\$-$ }& \makecell{ $ 0.4034714 - 
 0.0788704i$ \\
 $ 0.3763030 - 0.2406528i$\\
 $-$\\
 $-$} & \makecell{ $ 0.6108024- 0.0792061i $ \\ $ 0.5925156 - 0.2393481i $\\ $ 0.5559537 - 0.4050658i $\\ $-$ } &\makecell{$0.8170486 - 
0.0793308i$\\$0.8032845 - 
0.2389501i$\\$0.7757270 - 0.4015650i$\\ $0.7344624 - 0.5695741i$
} \\
 \hline
$ \xi_{\rm cr}$  &  \makecell{ $ 0 $ \\ $ 1 $\\ $ 2$\\ $3$ } &     \makecell{ $0.1920630 - 0.0769408 i$ \\ $0.141 - 0.249 i$\\ $-$\\$-$ }& \makecell{ $ 0.4038364-0.0784311i$ \\ $ 0.3761295 - 0.2395338i$\\
$0.3219361 - 0.4151015i$\\$-$ } & \makecell{ $0.6114677 - 0.0787573i $ \\ $ 0.5927929 - 0.2380928i $\\ $0.5555344 - 0.4032666i$\\ $0.5005594 - 0.5793100i$ } & \makecell{ $ 0.8179915 - 0.0788787 i$ \\ $ 0.8039287 - 0.2376456i $ \\ $0.7758066 - 0.3995604i$\\
$0.7337804 - 0.5671102i$} \\
   \hline

\end{tabular}
\end{table*}
\FloatBarrier
\bibliography{references.bib}

%apsrev4-2.bst 2019-01-14 (MD) hand-edited version of apsrev4-1.bst
%Control: key (0)
%Control: author (8) initials jnrlst
%Control: editor formatted (1) identically to author
%Control: production of article title (0) allowed
%Control: page (0) single
%Control: year (1) truncated
%Control: production of eprint (0) enabled
\begin{thebibliography}{97}%
\makeatletter
\providecommand \@ifxundefined [1]{%
 \@ifx{#1\undefined}
}%
\providecommand \@ifnum [1]{%
 \ifnum #1\expandafter \@firstoftwo
 \else \expandafter \@secondoftwo
 \fi
}%
\providecommand \@ifx [1]{%
 \ifx #1\expandafter \@firstoftwo
 \else \expandafter \@secondoftwo
 \fi
}%
\providecommand \natexlab [1]{#1}%
\providecommand \enquote  [1]{``#1''}%
\providecommand \bibnamefont  [1]{#1}%
\providecommand \bibfnamefont [1]{#1}%
\providecommand \citenamefont [1]{#1}%
\providecommand \href@noop [0]{\@secondoftwo}%
\providecommand \href [0]{\begingroup \@sanitize@url \@href}%
\providecommand \@href[1]{\@@startlink{#1}\@@href}%
\providecommand \@@href[1]{\endgroup#1\@@endlink}%
\providecommand \@sanitize@url [0]{\catcode `\\12\catcode `\$12\catcode `\&12\catcode `\#12\catcode `\^12\catcode `\_12\catcode `\%12\relax}%
\providecommand \@@startlink[1]{}%
\providecommand \@@endlink[0]{}%
\providecommand \url  [0]{\begingroup\@sanitize@url \@url }%
\providecommand \@url [1]{\endgroup\@href {#1}{\urlprefix }}%
\providecommand \urlprefix  [0]{URL }%
\providecommand \Eprint [0]{\href }%
\providecommand \doibase [0]{https://doi.org/}%
\providecommand \selectlanguage [0]{\@gobble}%
\providecommand \bibinfo  [0]{\@secondoftwo}%
\providecommand \bibfield  [0]{\@secondoftwo}%
\providecommand \translation [1]{[#1]}%
\providecommand \BibitemOpen [0]{}%
\providecommand \bibitemStop [0]{}%
\providecommand \bibitemNoStop [0]{.\EOS\space}%
\providecommand \EOS [0]{\spacefactor3000\relax}%
\providecommand \BibitemShut  [1]{\csname bibitem#1\endcsname}%
\let\auto@bib@innerbib\@empty
%</preamble>
\bibitem [{\citenamefont {Penrose}(1965)}]{PhysRevLett.14.57}%
  \BibitemOpen
  \bibfield  {author} {\bibinfo {author} {\bibfnamefont {R.}~\bibnamefont {Penrose}},\ }\bibfield  {title} {\bibinfo {title} {Gravitational collapse and space-time singularities},\ }\href {https://doi.org/10.1103/PhysRevLett.14.57} {\bibfield  {journal} {\bibinfo  {journal} {Phys. Rev. Lett.}\ }\textbf {\bibinfo {volume} {14}},\ \bibinfo {pages} {57} (\bibinfo {year} {1965})}\BibitemShut {NoStop}%
\bibitem [{\citenamefont {Hawking}\ and\ \citenamefont {Penrose}(1970)}]{Hawking:1970zqf}%
  \BibitemOpen
  \bibfield  {author} {\bibinfo {author} {\bibfnamefont {S.~W.}\ \bibnamefont {Hawking}}\ and\ \bibinfo {author} {\bibfnamefont {R.}~\bibnamefont {Penrose}},\ }\bibfield  {title} {\bibinfo {title} {{The Singularities of gravitational collapse and cosmology}},\ }\href {https://doi.org/10.1098/rspa.1970.0021} {\bibfield  {journal} {\bibinfo  {journal} {Proc. Roy. Soc. Lond. A}\ }\textbf {\bibinfo {volume} {314}},\ \bibinfo {pages} {529} (\bibinfo {year} {1970})}\BibitemShut {NoStop}%
\bibitem [{\citenamefont {Senovilla}(1998)}]{Senovilla_1998}%
  \BibitemOpen
  \bibfield  {author} {\bibinfo {author} {\bibfnamefont {J.~M.~M.}\ \bibnamefont {Senovilla}},\ }\bibfield  {title} {\bibinfo {title} {Singularity theorems and their consequences},\ }\href {https://doi.org/10.1023/a:1018801101244} {\bibfield  {journal} {\bibinfo  {journal} {General Relativity and Gravitation}\ }\textbf {\bibinfo {volume} {30}},\ \bibinfo {pages} {701–848} (\bibinfo {year} {1998})}\BibitemShut {NoStop}%
\bibitem [{\citenamefont {Penrose}(1969)}]{Penrose:1969pc}%
  \BibitemOpen
  \bibfield  {author} {\bibinfo {author} {\bibfnamefont {R.}~\bibnamefont {Penrose}},\ }\bibfield  {title} {\bibinfo {title} {{Gravitational collapse: The role of general relativity}},\ }\href {https://doi.org/10.1023/A:1016578408204} {\bibfield  {journal} {\bibinfo  {journal} {Riv. Nuovo Cim.}\ }\textbf {\bibinfo {volume} {1}},\ \bibinfo {pages} {252} (\bibinfo {year} {1969})}\BibitemShut {NoStop}%
\bibitem [{\citenamefont {{Bardeen}}(1968)}]{Bardeen}%
  \BibitemOpen
  \bibfield  {author} {\bibinfo {author} {\bibfnamefont {J.}~\bibnamefont {{Bardeen}}},\ }\bibfield  {title} {\bibinfo {title} {{Non-singular general relativistic gravitational collapse}},\ }in\ \href@noop {} {\emph {\bibinfo {booktitle} {Proceedings of the 5th International Conference on Gravitation and the Theory of Relativity}}}\ (\bibinfo {year} {1968})\ p.~\bibinfo {pages} {87}\BibitemShut {NoStop}%
\bibitem [{\citenamefont {Ay\'on-Beato}\ and\ \citenamefont {Garc\'{\i}a}(1998)}]{PhysRevLett.80.5056}%
  \BibitemOpen
  \bibfield  {author} {\bibinfo {author} {\bibfnamefont {E.}~\bibnamefont {Ay\'on-Beato}}\ and\ \bibinfo {author} {\bibfnamefont {A.}~\bibnamefont {Garc\'{\i}a}},\ }\bibfield  {title} {\bibinfo {title} {Regular black hole in general relativity coupled to nonlinear electrodynamics},\ }\href {https://doi.org/10.1103/PhysRevLett.80.5056} {\bibfield  {journal} {\bibinfo  {journal} {Phys. Rev. Lett.}\ }\textbf {\bibinfo {volume} {80}},\ \bibinfo {pages} {5056} (\bibinfo {year} {1998})}\BibitemShut {NoStop}%
\bibitem [{\citenamefont {{Dymnikova}}(1992)}]{1992GReGr..24..235D}%
  \BibitemOpen
  \bibfield  {author} {\bibinfo {author} {\bibfnamefont {I.}~\bibnamefont {{Dymnikova}}},\ }\bibfield  {title} {\bibinfo {title} {{Vacuum nonsingular black hole}},\ }\href {https://doi.org/10.1007/BF00760226} {\bibfield  {journal} {\bibinfo  {journal} {General Relativity and Gravitation}\ }\textbf {\bibinfo {volume} {24}},\ \bibinfo {pages} {235} (\bibinfo {year} {1992})}\BibitemShut {NoStop}%
\bibitem [{\citenamefont {Hayward}(2006)}]{PhysRevLett.96.031103}%
  \BibitemOpen
  \bibfield  {author} {\bibinfo {author} {\bibfnamefont {S.~A.}\ \bibnamefont {Hayward}},\ }\bibfield  {title} {\bibinfo {title} {Formation and evaporation of nonsingular black holes},\ }\href {https://doi.org/10.1103/PhysRevLett.96.031103} {\bibfield  {journal} {\bibinfo  {journal} {Phys. Rev. Lett.}\ }\textbf {\bibinfo {volume} {96}},\ \bibinfo {pages} {031103} (\bibinfo {year} {2006})}\BibitemShut {NoStop}%
\bibitem [{\citenamefont {{Bronnikov}}\ and\ \citenamefont {{Fabris}}(2006)}]{2006Bronnikov}%
  \BibitemOpen
  \bibfield  {author} {\bibinfo {author} {\bibfnamefont {K.~A.}\ \bibnamefont {{Bronnikov}}}\ and\ \bibinfo {author} {\bibfnamefont {J.~C.}\ \bibnamefont {{Fabris}}},\ }\bibfield  {title} {\bibinfo {title} {{Regular Phantom Black Holes}},\ }\href {https://doi.org/10.1103/PhysRevLett.96.251101} {\bibfield  {journal} {\bibinfo  {journal} {\prl}\ }\textbf {\bibinfo {volume} {96}},\ \bibinfo {eid} {251101} (\bibinfo {year} {2006})},\ \Eprint {https://arxiv.org/abs/gr-qc/0511109} {arXiv:gr-qc/0511109 [gr-qc]} \BibitemShut {NoStop}%
\bibitem [{\citenamefont {Maeda}(2022)}]{Maeda_2022}%
  \BibitemOpen
  \bibfield  {author} {\bibinfo {author} {\bibfnamefont {H.}~\bibnamefont {Maeda}},\ }\bibfield  {title} {\bibinfo {title} {Quest for realistic non-singular black-hole geometries: regular-center type},\ }\bibfield  {journal} {\bibinfo  {journal} {Journal of High Energy Physics}\ }\textbf {\bibinfo {volume} {2022}},\ \href {https://doi.org/10.1007/jhep11(2022)108} {10.1007/jhep11(2022)108} (\bibinfo {year} {2022})\BibitemShut {NoStop}%
\bibitem [{\citenamefont {{Burinskii}}\ \emph {et~al.}(2002)\citenamefont {{Burinskii}}, \citenamefont {{Elizalde}}, \citenamefont {{Hildebrandt}},\ and\ \citenamefont {{Magli}}}]{2002PhRvD..65f4039B}%
  \BibitemOpen
  \bibfield  {author} {\bibinfo {author} {\bibfnamefont {A.}~\bibnamefont {{Burinskii}}}, \bibinfo {author} {\bibfnamefont {E.}~\bibnamefont {{Elizalde}}}, \bibinfo {author} {\bibfnamefont {S.~R.}\ \bibnamefont {{Hildebrandt}}},\ and\ \bibinfo {author} {\bibfnamefont {G.}~\bibnamefont {{Magli}}},\ }\bibfield  {title} {\bibinfo {title} {{Regular sources of the Kerr-Schild class for rotating and nonrotating black hole solutions}},\ }\href {https://doi.org/10.1103/PhysRevD.65.064039} {\bibfield  {journal} {\bibinfo  {journal} {\prd}\ }\textbf {\bibinfo {volume} {65}},\ \bibinfo {eid} {064039} (\bibinfo {year} {2002})},\ \Eprint {https://arxiv.org/abs/gr-qc/0109085} {arXiv:gr-qc/0109085 [gr-qc]} \BibitemShut {NoStop}%
\bibitem [{\citenamefont {Fan}\ and\ \citenamefont {Wang}(2016)}]{Fan_2016}%
  \BibitemOpen
  \bibfield  {author} {\bibinfo {author} {\bibfnamefont {Z.-Y.}\ \bibnamefont {Fan}}\ and\ \bibinfo {author} {\bibfnamefont {X.}~\bibnamefont {Wang}},\ }\bibfield  {title} {\bibinfo {title} {Construction of regular black holes in general relativity},\ }\bibfield  {journal} {\bibinfo  {journal} {Physical Review D}\ }\textbf {\bibinfo {volume} {94}},\ \href {https://doi.org/10.1103/physrevd.94.124027} {10.1103/physrevd.94.124027} (\bibinfo {year} {2016})\BibitemShut {NoStop}%
\bibitem [{\citenamefont {Lobo}\ \emph {et~al.}(2021)\citenamefont {Lobo}, \citenamefont {Rodrigues}, \citenamefont {Silva}, \citenamefont {Simpson},\ and\ \citenamefont {Visser}}]{PhysRevD.103.084052}%
  \BibitemOpen
  \bibfield  {author} {\bibinfo {author} {\bibfnamefont {F.~S.~N.}\ \bibnamefont {Lobo}}, \bibinfo {author} {\bibfnamefont {M.~E.}\ \bibnamefont {Rodrigues}}, \bibinfo {author} {\bibfnamefont {M.~V. d.~S.}\ \bibnamefont {Silva}}, \bibinfo {author} {\bibfnamefont {A.}~\bibnamefont {Simpson}},\ and\ \bibinfo {author} {\bibfnamefont {M.}~\bibnamefont {Visser}},\ }\bibfield  {title} {\bibinfo {title} {Novel black-bounce spacetimes: Wormholes, regularity, energy conditions, and causal structure},\ }\href {https://doi.org/10.1103/PhysRevD.103.084052} {\bibfield  {journal} {\bibinfo  {journal} {Phys. Rev. D}\ }\textbf {\bibinfo {volume} {103}},\ \bibinfo {pages} {084052} (\bibinfo {year} {2021})}\BibitemShut {NoStop}%
\bibitem [{\citenamefont {{Giacchini}}\ and\ \citenamefont {{Netto}}(2023)}]{2023arXiv230712357G}%
  \BibitemOpen
  \bibfield  {author} {\bibinfo {author} {\bibfnamefont {B.~L.}\ \bibnamefont {{Giacchini}}}\ and\ \bibinfo {author} {\bibfnamefont {T.~d.~P.}\ \bibnamefont {{Netto}}},\ }\bibfield  {title} {\bibinfo {title} {{Regular black holes from higher-derivative effective delta sources}},\ }\href {https://doi.org/10.48550/arXiv.2307.12357} {\bibfield  {journal} {\bibinfo  {journal} {arXiv e-prints}\ ,\ \bibinfo {eid} {arXiv:2307.12357}} (\bibinfo {year} {2023})},\ \Eprint {https://arxiv.org/abs/2307.12357} {arXiv:2307.12357 [gr-qc]} \BibitemShut {NoStop}%
\bibitem [{\citenamefont {Ovalle}\ \emph {et~al.}(2023)\citenamefont {Ovalle}, \citenamefont {Casadio},\ and\ \citenamefont {Giusti}}]{Ovalle_2023}%
  \BibitemOpen
  \bibfield  {author} {\bibinfo {author} {\bibfnamefont {J.}~\bibnamefont {Ovalle}}, \bibinfo {author} {\bibfnamefont {R.}~\bibnamefont {Casadio}},\ and\ \bibinfo {author} {\bibfnamefont {A.}~\bibnamefont {Giusti}},\ }\bibfield  {title} {\bibinfo {title} {Regular hairy black holes through minkowski deformation},\ }\href {https://doi.org/10.1016/j.physletb.2023.138085} {\bibfield  {journal} {\bibinfo  {journal} {Physics Letters B}\ }\textbf {\bibinfo {volume} {844}},\ \bibinfo {pages} {138085} (\bibinfo {year} {2023})}\BibitemShut {NoStop}%
\bibitem [{\citenamefont {Bueno}\ \emph {et~al.}(2024)\citenamefont {Bueno}, \citenamefont {Cano},\ and\ \citenamefont {Hennigar}}]{Bueno:2024dgm}%
  \BibitemOpen
  \bibfield  {author} {\bibinfo {author} {\bibfnamefont {P.}~\bibnamefont {Bueno}}, \bibinfo {author} {\bibfnamefont {P.~A.}\ \bibnamefont {Cano}},\ and\ \bibinfo {author} {\bibfnamefont {R.~A.}\ \bibnamefont {Hennigar}},\ }\bibfield  {title} {\bibinfo {title} {{Regular Black Holes From Pure Gravity}},\ }\href@noop {} {\  (\bibinfo {year} {2024})},\ \Eprint {https://arxiv.org/abs/2403.04827} {arXiv:2403.04827 [gr-qc]} \BibitemShut {NoStop}%
\bibitem [{\citenamefont {{Junior}}\ \emph {et~al.}(2024)\citenamefont {{Junior}}, \citenamefont {{Lobo}},\ and\ \citenamefont {{Rodrigues}}}]{2024CQGra..41e5012J}%
  \BibitemOpen
  \bibfield  {author} {\bibinfo {author} {\bibfnamefont {J.~T. S.~S.}\ \bibnamefont {{Junior}}}, \bibinfo {author} {\bibfnamefont {F.~S.~N.}\ \bibnamefont {{Lobo}}},\ and\ \bibinfo {author} {\bibfnamefont {M.~E.}\ \bibnamefont {{Rodrigues}}},\ }\bibfield  {title} {\bibinfo {title} {{(Regular) Black holes in conformal Killing gravity coupled to nonlinear electrodynamics and scalar fields}},\ }\href {https://doi.org/10.1088/1361-6382/ad210e} {\bibfield  {journal} {\bibinfo  {journal} {Classical and Quantum Gravity}\ }\textbf {\bibinfo {volume} {41}},\ \bibinfo {eid} {055012} (\bibinfo {year} {2024})},\ \Eprint {https://arxiv.org/abs/2310.19508} {arXiv:2310.19508 [gr-qc]} \BibitemShut {NoStop}%
\bibitem [{\citenamefont {Carballo-Rubio}\ \emph {et~al.}(2022)\citenamefont {Carballo-Rubio}, \citenamefont {Di~Filippo}, \citenamefont {Liberati},\ and\ \citenamefont {Visser}}]{Carballo_Rubio_2022}%
  \BibitemOpen
  \bibfield  {author} {\bibinfo {author} {\bibfnamefont {R.}~\bibnamefont {Carballo-Rubio}}, \bibinfo {author} {\bibfnamefont {F.}~\bibnamefont {Di~Filippo}}, \bibinfo {author} {\bibfnamefont {S.}~\bibnamefont {Liberati}},\ and\ \bibinfo {author} {\bibfnamefont {M.}~\bibnamefont {Visser}},\ }\bibfield  {title} {\bibinfo {title} {Geodesically complete black holes in lorentz-violating gravity},\ }\bibfield  {journal} {\bibinfo  {journal} {Journal of High Energy Physics}\ }\textbf {\bibinfo {volume} {2022}},\ \href {https://doi.org/10.1007/jhep02(2022)122} {10.1007/jhep02(2022)122} (\bibinfo {year} {2022})\BibitemShut {NoStop}%
\bibitem [{\citenamefont {{Bakopoulos}}\ \emph {et~al.}(2024)\citenamefont {{Bakopoulos}}, \citenamefont {{Charmousis}}, \citenamefont {{Kanti}}, \citenamefont {{Lecoeur}},\ and\ \citenamefont {{Nakas}}}]{2024PhRvD.109b4032B}%
  \BibitemOpen
  \bibfield  {author} {\bibinfo {author} {\bibfnamefont {A.}~\bibnamefont {{Bakopoulos}}}, \bibinfo {author} {\bibfnamefont {C.}~\bibnamefont {{Charmousis}}}, \bibinfo {author} {\bibfnamefont {P.}~\bibnamefont {{Kanti}}}, \bibinfo {author} {\bibfnamefont {N.}~\bibnamefont {{Lecoeur}}},\ and\ \bibinfo {author} {\bibfnamefont {T.}~\bibnamefont {{Nakas}}},\ }\bibfield  {title} {\bibinfo {title} {{Black holes with primary scalar hair}},\ }\href {https://doi.org/10.1103/PhysRevD.109.024032} {\bibfield  {journal} {\bibinfo  {journal} {\prd}\ }\textbf {\bibinfo {volume} {109}},\ \bibinfo {eid} {024032} (\bibinfo {year} {2024})},\ \Eprint {https://arxiv.org/abs/2310.11919} {arXiv:2310.11919 [gr-qc]} \BibitemShut {NoStop}%
\bibitem [{\citenamefont {{Berej}}\ \emph {et~al.}(2006)\citenamefont {{Berej}}, \citenamefont {{Matyjasek}}, \citenamefont {{Tryniecki}},\ and\ \citenamefont {{Woronowicz}}}]{2006GReGr..38..885B}%
  \BibitemOpen
  \bibfield  {author} {\bibinfo {author} {\bibfnamefont {W.}~\bibnamefont {{Berej}}}, \bibinfo {author} {\bibfnamefont {J.}~\bibnamefont {{Matyjasek}}}, \bibinfo {author} {\bibfnamefont {D.}~\bibnamefont {{Tryniecki}}},\ and\ \bibinfo {author} {\bibfnamefont {M.}~\bibnamefont {{Woronowicz}}},\ }\bibfield  {title} {\bibinfo {title} {{Regular black holes in quadratic gravity}},\ }\href {https://doi.org/10.1007/s10714-006-0270-9} {\bibfield  {journal} {\bibinfo  {journal} {General Relativity and Gravitation}\ }\textbf {\bibinfo {volume} {38}},\ \bibinfo {pages} {885} (\bibinfo {year} {2006})},\ \Eprint {https://arxiv.org/abs/hep-th/0606185} {arXiv:hep-th/0606185 [hep-th]} \BibitemShut {NoStop}%
\bibitem [{\citenamefont {Mazza}\ and\ \citenamefont {Liberati}(2023)}]{Mazza_2023}%
  \BibitemOpen
  \bibfield  {author} {\bibinfo {author} {\bibfnamefont {J.}~\bibnamefont {Mazza}}\ and\ \bibinfo {author} {\bibfnamefont {S.}~\bibnamefont {Liberati}},\ }\bibfield  {title} {\bibinfo {title} {Regular black holes and horizonless ultra-compact objects in lorentz-violating gravity},\ }\bibfield  {journal} {\bibinfo  {journal} {Journal of High Energy Physics}\ }\textbf {\bibinfo {volume} {2023}},\ \href {https://doi.org/10.1007/jhep03(2023)199} {10.1007/jhep03(2023)199} (\bibinfo {year} {2023})\BibitemShut {NoStop}%
\bibitem [{\citenamefont {{Junior}}\ and\ \citenamefont {{Rodrigues}}(2023)}]{2023EPJC...83..475J}%
  \BibitemOpen
  \bibfield  {author} {\bibinfo {author} {\bibfnamefont {J.~T. S.~S.}\ \bibnamefont {{Junior}}}\ and\ \bibinfo {author} {\bibfnamefont {M.~E.}\ \bibnamefont {{Rodrigues}}},\ }\bibfield  {title} {\bibinfo {title} {{Coincident f (Q ) gravity: black holes, regular black holes, and black bounces}},\ }\href {https://doi.org/10.1140/epjc/s10052-023-11660-2} {\bibfield  {journal} {\bibinfo  {journal} {European Physical Journal C}\ }\textbf {\bibinfo {volume} {83}},\ \bibinfo {eid} {475} (\bibinfo {year} {2023})},\ \Eprint {https://arxiv.org/abs/2306.04661} {arXiv:2306.04661 [gr-qc]} \BibitemShut {NoStop}%
\bibitem [{\citenamefont {B\"ohmer}\ and\ \citenamefont {Vandersloot}(2007)}]{PhysRevD.76.104030}%
  \BibitemOpen
  \bibfield  {author} {\bibinfo {author} {\bibfnamefont {C.~G.}\ \bibnamefont {B\"ohmer}}\ and\ \bibinfo {author} {\bibfnamefont {K.}~\bibnamefont {Vandersloot}},\ }\bibfield  {title} {\bibinfo {title} {Loop quantum dynamics of the schwarzschild interior},\ }\href {https://doi.org/10.1103/PhysRevD.76.104030} {\bibfield  {journal} {\bibinfo  {journal} {Phys. Rev. D}\ }\textbf {\bibinfo {volume} {76}},\ \bibinfo {pages} {104030} (\bibinfo {year} {2007})}\BibitemShut {NoStop}%
\bibitem [{\citenamefont {{Modesto}}\ and\ \citenamefont {{Nicolini}}(2010)}]{2010PhRvD..82j4035M}%
  \BibitemOpen
  \bibfield  {author} {\bibinfo {author} {\bibfnamefont {L.}~\bibnamefont {{Modesto}}}\ and\ \bibinfo {author} {\bibfnamefont {P.}~\bibnamefont {{Nicolini}}},\ }\bibfield  {title} {\bibinfo {title} {{Charged rotating noncommutative black holes}},\ }\href {https://doi.org/10.1103/PhysRevD.82.104035} {\bibfield  {journal} {\bibinfo  {journal} {\prd}\ }\textbf {\bibinfo {volume} {82}},\ \bibinfo {eid} {104035} (\bibinfo {year} {2010})},\ \Eprint {https://arxiv.org/abs/1005.5605} {arXiv:1005.5605 [gr-qc]} \BibitemShut {NoStop}%
\bibitem [{\citenamefont {{Casadio}}\ \emph {et~al.}(2023)\citenamefont {{Casadio}}, \citenamefont {{Giusti}},\ and\ \citenamefont {{Ovalle}}}]{2023JHEP...05..118C}%
  \BibitemOpen
  \bibfield  {author} {\bibinfo {author} {\bibfnamefont {R.}~\bibnamefont {{Casadio}}}, \bibinfo {author} {\bibfnamefont {A.}~\bibnamefont {{Giusti}}},\ and\ \bibinfo {author} {\bibfnamefont {J.}~\bibnamefont {{Ovalle}}},\ }\bibfield  {title} {\bibinfo {title} {{Quantum rotating black holes}},\ }\href {https://doi.org/10.1007/JHEP05(2023)118} {\bibfield  {journal} {\bibinfo  {journal} {Journal of High Energy Physics}\ }\textbf {\bibinfo {volume} {2023}},\ \bibinfo {eid} {118} (\bibinfo {year} {2023})},\ \Eprint {https://arxiv.org/abs/2303.02713} {arXiv:2303.02713 [gr-qc]} \BibitemShut {NoStop}%
\bibitem [{\citenamefont {{Fernandes}}(2023)}]{2023PhRvD.108f1502F}%
  \BibitemOpen
  \bibfield  {author} {\bibinfo {author} {\bibfnamefont {P.~G.~S.}\ \bibnamefont {{Fernandes}}},\ }\bibfield  {title} {\bibinfo {title} {{Rotating black holes in semiclassical gravity}},\ }\href {https://doi.org/10.1103/PhysRevD.108.L061502} {\bibfield  {journal} {\bibinfo  {journal} {\prd}\ }\textbf {\bibinfo {volume} {108}},\ \bibinfo {eid} {L061502} (\bibinfo {year} {2023})},\ \Eprint {https://arxiv.org/abs/2305.10382} {arXiv:2305.10382 [gr-qc]} \BibitemShut {NoStop}%
\bibitem [{\citenamefont {{Nicolini}}\ \emph {et~al.}(2019)\citenamefont {{Nicolini}}, \citenamefont {{Spallucci}},\ and\ \citenamefont {{Wondrak}}}]{2019PhLB..79734888N}%
  \BibitemOpen
  \bibfield  {author} {\bibinfo {author} {\bibfnamefont {P.}~\bibnamefont {{Nicolini}}}, \bibinfo {author} {\bibfnamefont {E.}~\bibnamefont {{Spallucci}}},\ and\ \bibinfo {author} {\bibfnamefont {M.~F.}\ \bibnamefont {{Wondrak}}},\ }\bibfield  {title} {\bibinfo {title} {{Quantum corrected black holes from string T-duality}},\ }\href {https://doi.org/10.1016/j.physletb.2019.134888} {\bibfield  {journal} {\bibinfo  {journal} {Physics Letters B}\ }\textbf {\bibinfo {volume} {797}},\ \bibinfo {eid} {134888} (\bibinfo {year} {2019})},\ \Eprint {https://arxiv.org/abs/1902.11242} {arXiv:1902.11242 [gr-qc]} \BibitemShut {NoStop}%
\bibitem [{\citenamefont {Knorr}\ and\ \citenamefont {Platania}(2022)}]{Knorr_2022}%
  \BibitemOpen
  \bibfield  {author} {\bibinfo {author} {\bibfnamefont {B.}~\bibnamefont {Knorr}}\ and\ \bibinfo {author} {\bibfnamefont {A.}~\bibnamefont {Platania}},\ }\bibfield  {title} {\bibinfo {title} {Sifting quantum black holes through the principle of least action},\ }\bibfield  {journal} {\bibinfo  {journal} {Physical Review D}\ }\textbf {\bibinfo {volume} {106}},\ \href {https://doi.org/10.1103/physrevd.106.l021901} {10.1103/physrevd.106.l021901} (\bibinfo {year} {2022})\BibitemShut {NoStop}%
\bibitem [{\citenamefont {Platania}(2019)}]{Platania_2019}%
  \BibitemOpen
  \bibfield  {author} {\bibinfo {author} {\bibfnamefont {A.}~\bibnamefont {Platania}},\ }\bibfield  {title} {\bibinfo {title} {Dynamical renormalization of black-hole spacetimes},\ }\bibfield  {journal} {\bibinfo  {journal} {The European Physical Journal C}\ }\textbf {\bibinfo {volume} {79}},\ \href {https://doi.org/10.1140/epjc/s10052-019-6990-2} {10.1140/epjc/s10052-019-6990-2} (\bibinfo {year} {2019})\BibitemShut {NoStop}%
\bibitem [{\citenamefont {Brannlund}\ \emph {et~al.}(2009)\citenamefont {Brannlund}, \citenamefont {Kloster},\ and\ \citenamefont {DeBenedictis}}]{PhysRevD.79.084023}%
  \BibitemOpen
  \bibfield  {author} {\bibinfo {author} {\bibfnamefont {J.}~\bibnamefont {Brannlund}}, \bibinfo {author} {\bibfnamefont {S.}~\bibnamefont {Kloster}},\ and\ \bibinfo {author} {\bibfnamefont {A.}~\bibnamefont {DeBenedictis}},\ }\bibfield  {title} {\bibinfo {title} {Evolution of $\ensuremath{\Lambda}$ black holes in the minisuperspace approximation of loop quantum gravity},\ }\href {https://doi.org/10.1103/PhysRevD.79.084023} {\bibfield  {journal} {\bibinfo  {journal} {Phys. Rev. D}\ }\textbf {\bibinfo {volume} {79}},\ \bibinfo {pages} {084023} (\bibinfo {year} {2009})}\BibitemShut {NoStop}%
\bibitem [{\citenamefont {{Modesto}}(2010)}]{2010IJTP...49.1649M}%
  \BibitemOpen
  \bibfield  {author} {\bibinfo {author} {\bibfnamefont {L.}~\bibnamefont {{Modesto}}},\ }\bibfield  {title} {\bibinfo {title} {{Semiclassical Loop Quantum Black Hole}},\ }\href {https://doi.org/10.1007/s10773-010-0346-x} {\bibfield  {journal} {\bibinfo  {journal} {International Journal of Theoretical Physics}\ }\textbf {\bibinfo {volume} {49}},\ \bibinfo {pages} {1649} (\bibinfo {year} {2010})},\ \Eprint {https://arxiv.org/abs/0811.2196} {arXiv:0811.2196 [gr-qc]} \BibitemShut {NoStop}%
\bibitem [{\citenamefont {Koch}\ and\ \citenamefont {Saueressig}(2013)}]{Koch_2013}%
  \BibitemOpen
  \bibfield  {author} {\bibinfo {author} {\bibfnamefont {B.}~\bibnamefont {Koch}}\ and\ \bibinfo {author} {\bibfnamefont {F.}~\bibnamefont {Saueressig}},\ }\bibfield  {title} {\bibinfo {title} {Structural aspects of asymptotically safe black holes},\ }\href {https://doi.org/10.1088/0264-9381/31/1/015006} {\bibfield  {journal} {\bibinfo  {journal} {Classical and Quantum Gravity}\ }\textbf {\bibinfo {volume} {31}},\ \bibinfo {pages} {015006} (\bibinfo {year} {2013})}\BibitemShut {NoStop}%
\bibitem [{\citenamefont {Bonanno}\ and\ \citenamefont {Reuter}(2000{\natexlab{a}})}]{Bonanno_2000}%
  \BibitemOpen
  \bibfield  {author} {\bibinfo {author} {\bibfnamefont {A.}~\bibnamefont {Bonanno}}\ and\ \bibinfo {author} {\bibfnamefont {M.}~\bibnamefont {Reuter}},\ }\bibfield  {title} {\bibinfo {title} {Renormalization group improved black hole spacetimes},\ }\bibfield  {journal} {\bibinfo  {journal} {Physical Review D}\ }\textbf {\bibinfo {volume} {62}},\ \href {https://doi.org/10.1103/physrevd.62.043008} {10.1103/physrevd.62.043008} (\bibinfo {year} {2000}{\natexlab{a}})\BibitemShut {NoStop}%
\bibitem [{\citenamefont {Lewandowski}\ \emph {et~al.}(2023)\citenamefont {Lewandowski}, \citenamefont {Ma}, \citenamefont {Yang},\ and\ \citenamefont {Zhang}}]{PhysRevLett.130.101501}%
  \BibitemOpen
  \bibfield  {author} {\bibinfo {author} {\bibfnamefont {J.}~\bibnamefont {Lewandowski}}, \bibinfo {author} {\bibfnamefont {Y.}~\bibnamefont {Ma}}, \bibinfo {author} {\bibfnamefont {J.}~\bibnamefont {Yang}},\ and\ \bibinfo {author} {\bibfnamefont {C.}~\bibnamefont {Zhang}},\ }\bibfield  {title} {\bibinfo {title} {Quantum oppenheimer-snyder and swiss cheese models},\ }\href {https://doi.org/10.1103/PhysRevLett.130.101501} {\bibfield  {journal} {\bibinfo  {journal} {Phys. Rev. Lett.}\ }\textbf {\bibinfo {volume} {130}},\ \bibinfo {pages} {101501} (\bibinfo {year} {2023})}\BibitemShut {NoStop}%
\bibitem [{\citenamefont {{Bonanno}}\ \emph {et~al.}(2024)\citenamefont {{Bonanno}}, \citenamefont {{Malafarina}},\ and\ \citenamefont {{Panassiti}}}]{Bonanno_etal2024PhRvL}%
  \BibitemOpen
  \bibfield  {author} {\bibinfo {author} {\bibfnamefont {A.}~\bibnamefont {{Bonanno}}}, \bibinfo {author} {\bibfnamefont {D.}~\bibnamefont {{Malafarina}}},\ and\ \bibinfo {author} {\bibfnamefont {A.}~\bibnamefont {{Panassiti}}},\ }\bibfield  {title} {\bibinfo {title} {{Dust Collapse in Asymptotic Safety: A Path to Regular Black Holes}},\ }\href {https://doi.org/10.1103/PhysRevLett.132.031401} {\bibfield  {journal} {\bibinfo  {journal} {\prl}\ }\textbf {\bibinfo {volume} {132}},\ \bibinfo {eid} {031401} (\bibinfo {year} {2024})},\ \Eprint {https://arxiv.org/abs/2308.10890} {arXiv:2308.10890 [gr-qc]} \BibitemShut {NoStop}%
\bibitem [{\citenamefont {Binetti}\ \emph {et~al.}(2022)\citenamefont {Binetti}, \citenamefont {Del~Piano}, \citenamefont {Hohenegger}, \citenamefont {Pezzella},\ and\ \citenamefont {Sannino}}]{PhysRevD.106.046006}%
  \BibitemOpen
  \bibfield  {author} {\bibinfo {author} {\bibfnamefont {E.}~\bibnamefont {Binetti}}, \bibinfo {author} {\bibfnamefont {M.}~\bibnamefont {Del~Piano}}, \bibinfo {author} {\bibfnamefont {S.}~\bibnamefont {Hohenegger}}, \bibinfo {author} {\bibfnamefont {F.}~\bibnamefont {Pezzella}},\ and\ \bibinfo {author} {\bibfnamefont {F.}~\bibnamefont {Sannino}},\ }\bibfield  {title} {\bibinfo {title} {Effective theory of quantum black holes},\ }\href {https://doi.org/10.1103/PhysRevD.106.046006} {\bibfield  {journal} {\bibinfo  {journal} {Phys. Rev. D}\ }\textbf {\bibinfo {volume} {106}},\ \bibinfo {pages} {046006} (\bibinfo {year} {2022})}\BibitemShut {NoStop}%
\bibitem [{\citenamefont {Akil}\ \emph {et~al.}(2023)\citenamefont {Akil}, \citenamefont {Cadoni}, \citenamefont {Modesto}, \citenamefont {Oi},\ and\ \citenamefont {Sanna}}]{PhysRevD.108.044051}%
  \BibitemOpen
  \bibfield  {author} {\bibinfo {author} {\bibfnamefont {A.}~\bibnamefont {Akil}}, \bibinfo {author} {\bibfnamefont {M.}~\bibnamefont {Cadoni}}, \bibinfo {author} {\bibfnamefont {L.}~\bibnamefont {Modesto}}, \bibinfo {author} {\bibfnamefont {M.}~\bibnamefont {Oi}},\ and\ \bibinfo {author} {\bibfnamefont {A.~P.}\ \bibnamefont {Sanna}},\ }\bibfield  {title} {\bibinfo {title} {Semiclassical spacetimes at super-planckian scales from delocalized sources},\ }\href {https://doi.org/10.1103/PhysRevD.108.044051} {\bibfield  {journal} {\bibinfo  {journal} {Phys. Rev. D}\ }\textbf {\bibinfo {volume} {108}},\ \bibinfo {pages} {044051} (\bibinfo {year} {2023})}\BibitemShut {NoStop}%
\bibitem [{\citenamefont {{Carballo-Rubio}}\ \emph {et~al.}(2020)\citenamefont {{Carballo-Rubio}}, \citenamefont {{Di Filippo}}, \citenamefont {{Liberati}},\ and\ \citenamefont {{Visser}}}]{2020PhRvD.101h4047C}%
  \BibitemOpen
  \bibfield  {author} {\bibinfo {author} {\bibfnamefont {R.}~\bibnamefont {{Carballo-Rubio}}}, \bibinfo {author} {\bibfnamefont {F.}~\bibnamefont {{Di Filippo}}}, \bibinfo {author} {\bibfnamefont {S.}~\bibnamefont {{Liberati}}},\ and\ \bibinfo {author} {\bibfnamefont {M.}~\bibnamefont {{Visser}}},\ }\bibfield  {title} {\bibinfo {title} {{Geodesically complete black holes}},\ }\href {https://doi.org/10.1103/PhysRevD.101.084047} {\bibfield  {journal} {\bibinfo  {journal} {\prd}\ }\textbf {\bibinfo {volume} {101}},\ \bibinfo {eid} {084047} (\bibinfo {year} {2020})},\ \Eprint {https://arxiv.org/abs/1911.11200} {arXiv:1911.11200 [gr-qc]} \BibitemShut {NoStop}%
\bibitem [{\citenamefont {{Eichhorn}}(2012)}]{Eichhorn2012PhRvD}%
  \BibitemOpen
  \bibfield  {author} {\bibinfo {author} {\bibfnamefont {A.}~\bibnamefont {{Eichhorn}}},\ }\bibfield  {title} {\bibinfo {title} {{Quantum-gravity-induced matter self-interactions in the asymptotic-safety scenario}},\ }\href {https://doi.org/10.1103/PhysRevD.86.105021} {\bibfield  {journal} {\bibinfo  {journal} {\prd}\ }\textbf {\bibinfo {volume} {86}},\ \bibinfo {eid} {105021} (\bibinfo {year} {2012})},\ \Eprint {https://arxiv.org/abs/1204.0965} {arXiv:1204.0965 [gr-qc]} \BibitemShut {NoStop}%
\bibitem [{\citenamefont {Niedermaier}(2007)}]{Niedermaier_2007}%
  \BibitemOpen
  \bibfield  {author} {\bibinfo {author} {\bibfnamefont {M.}~\bibnamefont {Niedermaier}},\ }\bibfield  {title} {\bibinfo {title} {The asymptotic safety scenario in quantum gravity: an introduction},\ }\href {https://doi.org/10.1088/0264-9381/24/18/r01} {\bibfield  {journal} {\bibinfo  {journal} {Classical and Quantum Gravity}\ }\textbf {\bibinfo {volume} {24}},\ \bibinfo {pages} {R171–R230} (\bibinfo {year} {2007})}\BibitemShut {NoStop}%
\bibitem [{\citenamefont {Platania}(2023)}]{Platania_2023}%
  \BibitemOpen
  \bibfield  {author} {\bibinfo {author} {\bibfnamefont {A.}~\bibnamefont {Platania}},\ }\bibinfo {title} {Black holes in asymptotically safe gravity},\ in\ \href {https://doi.org/10.1007/978-981-19-3079-9_24-1} {\emph {\bibinfo {booktitle} {Handbook of Quantum Gravity}}}\ (\bibinfo  {publisher} {Springer Nature Singapore},\ \bibinfo {year} {2023})\ p.\ \bibinfo {pages} {1–65}\BibitemShut {NoStop}%
\bibitem [{\citenamefont {Del~Piano}\ \emph {et~al.}(2024)\citenamefont {Del~Piano}, \citenamefont {Hohenegger},\ and\ \citenamefont {Sannino}}]{PhysRevD.109.024045}%
  \BibitemOpen
  \bibfield  {author} {\bibinfo {author} {\bibfnamefont {M.}~\bibnamefont {Del~Piano}}, \bibinfo {author} {\bibfnamefont {S.}~\bibnamefont {Hohenegger}},\ and\ \bibinfo {author} {\bibfnamefont {F.}~\bibnamefont {Sannino}},\ }\bibfield  {title} {\bibinfo {title} {Quantum black hole physics from the event horizon},\ }\href {https://doi.org/10.1103/PhysRevD.109.024045} {\bibfield  {journal} {\bibinfo  {journal} {Phys. Rev. D}\ }\textbf {\bibinfo {volume} {109}},\ \bibinfo {pages} {024045} (\bibinfo {year} {2024})}\BibitemShut {NoStop}%
\bibitem [{\citenamefont {Borissova}\ and\ \citenamefont {Platania}(2023)}]{Borissova_2023}%
  \BibitemOpen
  \bibfield  {author} {\bibinfo {author} {\bibfnamefont {J.~N.}\ \bibnamefont {Borissova}}\ and\ \bibinfo {author} {\bibfnamefont {A.}~\bibnamefont {Platania}},\ }\bibfield  {title} {\bibinfo {title} {Formation and evaporation of quantum black holes from the decoupling mechanism in quantum gravity},\ }\bibfield  {journal} {\bibinfo  {journal} {Journal of High Energy Physics}\ }\textbf {\bibinfo {volume} {2023}},\ \href {https://doi.org/10.1007/jhep03(2023)046} {10.1007/jhep03(2023)046} (\bibinfo {year} {2023})\BibitemShut {NoStop}%
\bibitem [{\citenamefont {Pawlowski}\ and\ \citenamefont {Tränkle}(2023)}]{pawlowski2023effectiveactionblackhole}%
  \BibitemOpen
  \bibfield  {author} {\bibinfo {author} {\bibfnamefont {J.~M.}\ \bibnamefont {Pawlowski}}\ and\ \bibinfo {author} {\bibfnamefont {J.}~\bibnamefont {Tränkle}},\ }\href {https://arxiv.org/abs/2309.17043} {\bibinfo {title} {Effective action and black hole solutions in asymptotically safe quantum gravity}} (\bibinfo {year} {2023}),\ \Eprint {https://arxiv.org/abs/2309.17043} {arXiv:2309.17043 [hep-th]} \BibitemShut {NoStop}%
\bibitem [{\citenamefont {{B.~P.~Abbott {\textit{et al}}}}(2016)}]{LIGO+VIRGO2016}%
  \BibitemOpen
  \bibfield  {author} {\bibinfo {author} {\bibnamefont {{B.~P.~Abbott {\textit{et al}}}}} (\bibinfo {collaboration} {LIGO Scientific Collaboration and Virgo Collaboration}),\ }\bibfield  {title} {\bibinfo {title} {Observation of gravitational waves from a binary black hole merger},\ }\href {https://doi.org/10.1103/PhysRevLett.116.061102} {\bibfield  {journal} {\bibinfo  {journal} {Phys. Rev. Lett.}\ }\textbf {\bibinfo {volume} {116}},\ \bibinfo {pages} {061102} (\bibinfo {year} {2016})}\BibitemShut {NoStop}%
\bibitem [{\citenamefont {{R.~Abbott {\textit{et al}}}}(2023)}]{LIGO-VIRGO-KAGRA2023}%
  \BibitemOpen
  \bibfield  {author} {\bibinfo {author} {\bibnamefont {{R.~Abbott {\textit{et al}}}}} (\bibinfo {collaboration} {Ligo Scientific Collaboration, VIRGO Collaboration, Kagra Collaboration}),\ }\bibfield  {title} {\bibinfo {title} {{GWTC-3: Compact Binary Coalescences Observed by LIGO and Virgo during the Second Part of the Third Observing Run}},\ }\href {https://doi.org/10.1103/PhysRevX.13.041039} {\bibfield  {journal} {\bibinfo  {journal} {Physical Review X}\ }\textbf {\bibinfo {volume} {13}},\ \bibinfo {eid} {041039} (\bibinfo {year} {2023})},\ \Eprint {https://arxiv.org/abs/2111.03606} {arXiv:2111.03606 [gr-qc]} \BibitemShut {NoStop}%
\bibitem [{\citenamefont {{ K. {Akiyama} and \textit{et al}}}(2019)}]{Akiyama2019}%
  \BibitemOpen
  \bibfield  {author} {\bibinfo {author} {\bibnamefont {{ K. {Akiyama} and \textit{et al}}}} (\bibinfo {collaboration} {Event Horizon Telescope Collaboration}),\ }\bibfield  {title} {\bibinfo {title} {{First M87 Event Horizon Telescope Results. I. The Shadow of the Supermassive Black Hole}},\ }\href {https://doi.org/10.3847/2041-8213/ab0ec7} {\bibfield  {journal} {\bibinfo  {journal} {The Astrophysical Journal Letters}\ }\textbf {\bibinfo {volume} {875}},\ \bibinfo {eid} {L1} (\bibinfo {year} {2019})}\BibitemShut {NoStop}%
\bibitem [{\citenamefont {{ K. {Akiyama} and \textit{et al}}}(2022)}]{EHT2022}%
  \BibitemOpen
  \bibfield  {author} {\bibinfo {author} {\bibnamefont {{ K. {Akiyama} and \textit{et al}}}} (\bibinfo {collaboration} {Event Horizon Telescope Collaboration}),\ }\bibfield  {title} {\bibinfo {title} {{First Sagittarius A* Event Horizon Telescope Results. I. The Shadow of the Supermassive Black Hole in the Center of the Milky Way}},\ }\href {https://doi.org/10.3847/2041-8213/ac6674} {\bibfield  {journal} {\bibinfo  {journal} {The Astrophysical Journal Letters}\ }\textbf {\bibinfo {volume} {930}},\ \bibinfo {eid} {L12} (\bibinfo {year} {2022})}\BibitemShut {NoStop}%
\bibitem [{\citenamefont {{ K.~G. Arun \textit{et al}}}(2022)}]{LISA2022LRR}%
  \BibitemOpen
  \bibfield  {author} {\bibinfo {author} {\bibnamefont {{ K.~G. Arun \textit{et al}}}} (\bibinfo {collaboration} {LISA Consortium}),\ }\bibfield  {title} {\bibinfo {title} {{New horizons for fundamental physics with LISA}},\ }\href {https://doi.org/10.1007/s41114-022-00036-9} {\bibfield  {journal} {\bibinfo  {journal} {Living Reviews in Relativity}\ }\textbf {\bibinfo {volume} {25}},\ \bibinfo {eid} {4} (\bibinfo {year} {2022})},\ \Eprint {https://arxiv.org/abs/2205.01597} {arXiv:2205.01597 [gr-qc]} \BibitemShut {NoStop}%
\bibitem [{\citenamefont {Abdujabbarov}\ \emph {et~al.}(2016)\citenamefont {Abdujabbarov}, \citenamefont {Amir}, \citenamefont {Ahmedov},\ and\ \citenamefont {Ghosh}}]{Abdujabbarov_2016}%
  \BibitemOpen
  \bibfield  {author} {\bibinfo {author} {\bibfnamefont {A.}~\bibnamefont {Abdujabbarov}}, \bibinfo {author} {\bibfnamefont {M.}~\bibnamefont {Amir}}, \bibinfo {author} {\bibfnamefont {B.}~\bibnamefont {Ahmedov}},\ and\ \bibinfo {author} {\bibfnamefont {S.~G.}\ \bibnamefont {Ghosh}},\ }\bibfield  {title} {\bibinfo {title} {Shadow of rotating regular black holes},\ }\bibfield  {journal} {\bibinfo  {journal} {Physical Review D}\ }\textbf {\bibinfo {volume} {93}},\ \href {https://doi.org/10.1103/physrevd.93.104004} {10.1103/physrevd.93.104004} (\bibinfo {year} {2016})\BibitemShut {NoStop}%
\bibitem [{\citenamefont {Ling}\ and\ \citenamefont {Wu}(2022)}]{Ling_2022}%
  \BibitemOpen
  \bibfield  {author} {\bibinfo {author} {\bibfnamefont {Y.}~\bibnamefont {Ling}}\ and\ \bibinfo {author} {\bibfnamefont {M.-H.}\ \bibnamefont {Wu}},\ }\bibfield  {title} {\bibinfo {title} {The shadows of regular black holes with asymptotic minkowski cores},\ }\href {https://doi.org/10.3390/sym14112415} {\bibfield  {journal} {\bibinfo  {journal} {Symmetry}\ }\textbf {\bibinfo {volume} {14}},\ \bibinfo {pages} {2415} (\bibinfo {year} {2022})}\BibitemShut {NoStop}%
\bibitem [{\citenamefont {Ghosh}\ \emph {et~al.}(2020)\citenamefont {Ghosh}, \citenamefont {Amir},\ and\ \citenamefont {Maharaj}}]{Ghosh_2020}%
  \BibitemOpen
  \bibfield  {author} {\bibinfo {author} {\bibfnamefont {S.~G.}\ \bibnamefont {Ghosh}}, \bibinfo {author} {\bibfnamefont {M.}~\bibnamefont {Amir}},\ and\ \bibinfo {author} {\bibfnamefont {S.~D.}\ \bibnamefont {Maharaj}},\ }\bibfield  {title} {\bibinfo {title} {Ergosphere and shadow of a rotating regular black hole},\ }\href {https://doi.org/10.1016/j.nuclphysb.2020.115088} {\bibfield  {journal} {\bibinfo  {journal} {Nuclear Physics B}\ }\textbf {\bibinfo {volume} {957}},\ \bibinfo {pages} {115088} (\bibinfo {year} {2020})}\BibitemShut {NoStop}%
\bibitem [{\citenamefont {Sau}\ and\ \citenamefont {Moffat}(2023)}]{PhysRevD.107.124003}%
  \BibitemOpen
  \bibfield  {author} {\bibinfo {author} {\bibfnamefont {S.}~\bibnamefont {Sau}}\ and\ \bibinfo {author} {\bibfnamefont {J.~W.}\ \bibnamefont {Moffat}},\ }\bibfield  {title} {\bibinfo {title} {Shadow of a regular black hole in scalar-tensor-vector gravity theory},\ }\href {https://doi.org/10.1103/PhysRevD.107.124003} {\bibfield  {journal} {\bibinfo  {journal} {Phys. Rev. D}\ }\textbf {\bibinfo {volume} {107}},\ \bibinfo {pages} {124003} (\bibinfo {year} {2023})}\BibitemShut {NoStop}%
\bibitem [{\citenamefont {Dymnikova}\ and\ \citenamefont {Kraav}(2019)}]{universe5070163}%
  \BibitemOpen
  \bibfield  {author} {\bibinfo {author} {\bibfnamefont {I.}~\bibnamefont {Dymnikova}}\ and\ \bibinfo {author} {\bibfnamefont {K.}~\bibnamefont {Kraav}},\ }\bibfield  {title} {\bibinfo {title} {Identification of a regular black hole by its shadow},\ }\bibfield  {journal} {\bibinfo  {journal} {Universe}\ }\textbf {\bibinfo {volume} {5}},\ \href {https://doi.org/10.3390/universe5070163} {10.3390/universe5070163} (\bibinfo {year} {2019})\BibitemShut {NoStop}%
\bibitem [{\citenamefont {Yang}\ \emph {et~al.}(2023)\citenamefont {Yang}, \citenamefont {Zhang},\ and\ \citenamefont {Ma}}]{Yang_2023}%
  \BibitemOpen
  \bibfield  {author} {\bibinfo {author} {\bibfnamefont {J.}~\bibnamefont {Yang}}, \bibinfo {author} {\bibfnamefont {C.}~\bibnamefont {Zhang}},\ and\ \bibinfo {author} {\bibfnamefont {Y.}~\bibnamefont {Ma}},\ }\bibfield  {title} {\bibinfo {title} {Shadow and stability of quantum-corrected black holes},\ }\bibfield  {journal} {\bibinfo  {journal} {The European Physical Journal C}\ }\textbf {\bibinfo {volume} {83}},\ \href {https://doi.org/10.1140/epjc/s10052-023-11800-8} {10.1140/epjc/s10052-023-11800-8} (\bibinfo {year} {2023})\BibitemShut {NoStop}%
\bibitem [{\citenamefont {{Walia}}\ \emph {et~al.}(2022)\citenamefont {{Walia}}, \citenamefont {{Ghosh}},\ and\ \citenamefont {{Maharaj}}}]{2022ApJ...939...77W}%
  \BibitemOpen
  \bibfield  {author} {\bibinfo {author} {\bibfnamefont {R.~K.}\ \bibnamefont {{Walia}}}, \bibinfo {author} {\bibfnamefont {S.~G.}\ \bibnamefont {{Ghosh}}},\ and\ \bibinfo {author} {\bibfnamefont {S.~D.}\ \bibnamefont {{Maharaj}}},\ }\bibfield  {title} {\bibinfo {title} {{Testing Rotating Regular Metrics with EHT Results of Sgr A*}},\ }\href {https://doi.org/10.3847/1538-4357/ac9623} {\bibfield  {journal} {\bibinfo  {journal} {\apj}\ }\textbf {\bibinfo {volume} {939}},\ \bibinfo {eid} {77} (\bibinfo {year} {2022})},\ \Eprint {https://arxiv.org/abs/2207.00078} {arXiv:2207.00078 [gr-qc]} \BibitemShut {NoStop}%
\bibitem [{\citenamefont {Flachi}\ and\ \citenamefont {Lemos}(2013)}]{Flachi_2013}%
  \BibitemOpen
  \bibfield  {author} {\bibinfo {author} {\bibfnamefont {A.}~\bibnamefont {Flachi}}\ and\ \bibinfo {author} {\bibfnamefont {J.~P.~S.}\ \bibnamefont {Lemos}},\ }\bibfield  {title} {\bibinfo {title} {Quasinormal modes of regular black holes},\ }\bibfield  {journal} {\bibinfo  {journal} {Physical Review D}\ }\textbf {\bibinfo {volume} {87}},\ \href {https://doi.org/10.1103/physrevd.87.024034} {10.1103/physrevd.87.024034} (\bibinfo {year} {2013})\BibitemShut {NoStop}%
\bibitem [{\citenamefont {{Konoplya}}\ \emph {et~al.}(2023)\citenamefont {{Konoplya}}, \citenamefont {{Stuchl{\'\i}k}}, \citenamefont {{Zhidenko}},\ and\ \citenamefont {{Zinhailo}}}]{2023PhRvD.107j4050K}%
  \BibitemOpen
  \bibfield  {author} {\bibinfo {author} {\bibfnamefont {R.~A.}\ \bibnamefont {{Konoplya}}}, \bibinfo {author} {\bibfnamefont {Z.}~\bibnamefont {{Stuchl{\'\i}k}}}, \bibinfo {author} {\bibfnamefont {A.}~\bibnamefont {{Zhidenko}}},\ and\ \bibinfo {author} {\bibfnamefont {A.~F.}\ \bibnamefont {{Zinhailo}}},\ }\bibfield  {title} {\bibinfo {title} {{Quasinormal modes of renormalization group improved Dymnikova regular black holes}},\ }\href {https://doi.org/10.1103/PhysRevD.107.104050} {\bibfield  {journal} {\bibinfo  {journal} {\prd}\ }\textbf {\bibinfo {volume} {107}},\ \bibinfo {eid} {104050} (\bibinfo {year} {2023})},\ \Eprint {https://arxiv.org/abs/2303.01987} {arXiv:2303.01987 [gr-qc]} \BibitemShut {NoStop}%
\bibitem [{\citenamefont {Gingrich}(2024)}]{PhysRevD.109.044044}%
  \BibitemOpen
  \bibfield  {author} {\bibinfo {author} {\bibfnamefont {D.~M.}\ \bibnamefont {Gingrich}},\ }\bibfield  {title} {\bibinfo {title} {Quasinormal modes of loop quantum black holes near the planck scale},\ }\href {https://doi.org/10.1103/PhysRevD.109.044044} {\bibfield  {journal} {\bibinfo  {journal} {Phys. Rev. D}\ }\textbf {\bibinfo {volume} {109}},\ \bibinfo {pages} {044044} (\bibinfo {year} {2024})}\BibitemShut {NoStop}%
\bibitem [{\citenamefont {{Simpson}}(2021)}]{2021Univ....7..418S}%
  \BibitemOpen
  \bibfield  {author} {\bibinfo {author} {\bibfnamefont {A.~M.}\ \bibnamefont {{Simpson}}},\ }\bibfield  {title} {\bibinfo {title} {{Ringing of the Regular Black Hole with Asymptotically Minkowski Core}},\ }\href {https://doi.org/10.3390/universe7110418} {\bibfield  {journal} {\bibinfo  {journal} {Universe}\ }\textbf {\bibinfo {volume} {7}},\ \bibinfo {eid} {418} (\bibinfo {year} {2021})},\ \Eprint {https://arxiv.org/abs/2109.11878} {arXiv:2109.11878 [gr-qc]} \BibitemShut {NoStop}%
\bibitem [{\citenamefont {{Konoplya}}\ \emph {et~al.}(2022)\citenamefont {{Konoplya}}, \citenamefont {{Zinhailo}}, \citenamefont {{Kunz}}, \citenamefont {{Stuchl{\'\i}k}},\ and\ \citenamefont {{Zhidenko}}}]{2022JCAP...10..091K}%
  \BibitemOpen
  \bibfield  {author} {\bibinfo {author} {\bibfnamefont {R.~A.}\ \bibnamefont {{Konoplya}}}, \bibinfo {author} {\bibfnamefont {A.~F.}\ \bibnamefont {{Zinhailo}}}, \bibinfo {author} {\bibfnamefont {J.}~\bibnamefont {{Kunz}}}, \bibinfo {author} {\bibfnamefont {Z.}~\bibnamefont {{Stuchl{\'\i}k}}},\ and\ \bibinfo {author} {\bibfnamefont {A.}~\bibnamefont {{Zhidenko}}},\ }\bibfield  {title} {\bibinfo {title} {{Quasinormal ringing of regular black holes in asymptotically safe gravity: the importance of overtones}},\ }\href {https://doi.org/10.1088/1475-7516/2022/10/091} {\bibfield  {journal} {\bibinfo  {journal} {JCAP}\ }\textbf {\bibinfo {volume} {2022}}\bibfield  {number} {\bibinfo  {number} { (10)},\ \bibinfo {eid} {091}},\ }\Eprint {https://arxiv.org/abs/2206.14714} {arXiv:2206.14714 [gr-qc]} \BibitemShut {NoStop}%
\bibitem [{\citenamefont {{Mahdavian Yekta}}\ \emph {et~al.}(2021)\citenamefont {{Mahdavian Yekta}}, \citenamefont {{Karimabadi}},\ and\ \citenamefont {{Alavi}}}]{2021AnPhy.43468603M}%
  \BibitemOpen
  \bibfield  {author} {\bibinfo {author} {\bibfnamefont {D.}~\bibnamefont {{Mahdavian Yekta}}}, \bibinfo {author} {\bibfnamefont {M.}~\bibnamefont {{Karimabadi}}},\ and\ \bibinfo {author} {\bibfnamefont {S.~A.}\ \bibnamefont {{Alavi}}},\ }\bibfield  {title} {\bibinfo {title} {{Quasinormal modes for non-minimally coupled scalar fields in regular black hole spacetimes: Grey-body factors, area spectrum and shadow radius}},\ }\href {https://doi.org/10.1016/j.aop.2021.168603} {\bibfield  {journal} {\bibinfo  {journal} {Annals of Physics}\ }\textbf {\bibinfo {volume} {434}},\ \bibinfo {eid} {168603} (\bibinfo {year} {2021})},\ \Eprint {https://arxiv.org/abs/1912.12017} {arXiv:1912.12017 [hep-th]} \BibitemShut {NoStop}%
\bibitem [{\citenamefont {{Meng}}\ and\ \citenamefont {{Zhang}}(2023)}]{2023CQGra..40s5024M}%
  \BibitemOpen
  \bibfield  {author} {\bibinfo {author} {\bibfnamefont {K.}~\bibnamefont {{Meng}}}\ and\ \bibinfo {author} {\bibfnamefont {S.-J.}\ \bibnamefont {{Zhang}}},\ }\bibfield  {title} {\bibinfo {title} {{Gravito-electromagnetic perturbations and QNMs of regular black holes}},\ }\href {https://doi.org/10.1088/1361-6382/acf3c6} {\bibfield  {journal} {\bibinfo  {journal} {Classical and Quantum Gravity}\ }\textbf {\bibinfo {volume} {40}},\ \bibinfo {eid} {195024} (\bibinfo {year} {2023})},\ \Eprint {https://arxiv.org/abs/2210.00295} {arXiv:2210.00295 [gr-qc]} \BibitemShut {NoStop}%
\bibitem [{\citenamefont {{Franzin}}\ \emph {et~al.}(2024)\citenamefont {{Franzin}}, \citenamefont {{Liberati}},\ and\ \citenamefont {{Vellucci}}}]{2024JCAP...01..020F}%
  \BibitemOpen
  \bibfield  {author} {\bibinfo {author} {\bibfnamefont {E.}~\bibnamefont {{Franzin}}}, \bibinfo {author} {\bibfnamefont {S.}~\bibnamefont {{Liberati}}},\ and\ \bibinfo {author} {\bibfnamefont {V.}~\bibnamefont {{Vellucci}}},\ }\bibfield  {title} {\bibinfo {title} {{From regular black holes to horizonless objects: quasi-normal modes, instabilities and spectroscopy}},\ }\href {https://doi.org/10.1088/1475-7516/2024/01/020} {\bibfield  {journal} {\bibinfo  {journal} {JCAP}\ }\textbf {\bibinfo {volume} {2024}}\bibfield  {number} {\bibinfo  {number} { (1)},\ \bibinfo {eid} {020}},\ }\Eprint {https://arxiv.org/abs/2310.11990} {arXiv:2310.11990 [gr-qc]} \BibitemShut {NoStop}%
\bibitem [{\citenamefont {Konoplya}\ \emph {et~al.}(2023)\citenamefont {Konoplya}, \citenamefont {Ovchinnikov},\ and\ \citenamefont {Ahmedov}}]{PhysRevD.108.104054}%
  \BibitemOpen
  \bibfield  {author} {\bibinfo {author} {\bibfnamefont {R.~A.}\ \bibnamefont {Konoplya}}, \bibinfo {author} {\bibfnamefont {D.}~\bibnamefont {Ovchinnikov}},\ and\ \bibinfo {author} {\bibfnamefont {B.}~\bibnamefont {Ahmedov}},\ }\bibfield  {title} {\bibinfo {title} {Bardeen spacetime as a quantum corrected schwarzschild black hole: Quasinormal modes and hawking radiation},\ }\href {https://doi.org/10.1103/PhysRevD.108.104054} {\bibfield  {journal} {\bibinfo  {journal} {Phys. Rev. D}\ }\textbf {\bibinfo {volume} {108}},\ \bibinfo {pages} {104054} (\bibinfo {year} {2023})}\BibitemShut {NoStop}%
\bibitem [{\citenamefont {Konoplya}\ and\ \citenamefont {Zhidenko}(2024)}]{PhysRevD.109.104005}%
  \BibitemOpen
  \bibfield  {author} {\bibinfo {author} {\bibfnamefont {R.~A.}\ \bibnamefont {Konoplya}}\ and\ \bibinfo {author} {\bibfnamefont {A.}~\bibnamefont {Zhidenko}},\ }\bibfield  {title} {\bibinfo {title} {Infinite tower of higher-curvature corrections: Quasinormal modes and late-time behavior of $d$-dimensional regular black holes},\ }\href {https://doi.org/10.1103/PhysRevD.109.104005} {\bibfield  {journal} {\bibinfo  {journal} {Phys. Rev. D}\ }\textbf {\bibinfo {volume} {109}},\ \bibinfo {pages} {104005} (\bibinfo {year} {2024})}\BibitemShut {NoStop}%
\bibitem [{\citenamefont {Zhang}\ \emph {et~al.}(2024)\citenamefont {Zhang}, \citenamefont {Gong}, \citenamefont {Fu}, \citenamefont {Wu},\ and\ \citenamefont {Pan}}]{zhang2024quasinormal}%
  \BibitemOpen
  \bibfield  {author} {\bibinfo {author} {\bibfnamefont {D.}~\bibnamefont {Zhang}}, \bibinfo {author} {\bibfnamefont {H.}~\bibnamefont {Gong}}, \bibinfo {author} {\bibfnamefont {G.}~\bibnamefont {Fu}}, \bibinfo {author} {\bibfnamefont {J.-P.}\ \bibnamefont {Wu}},\ and\ \bibinfo {author} {\bibfnamefont {Q.}~\bibnamefont {Pan}},\ }\href@noop {} {\bibinfo {title} {Quasinormal modes of a regular black hole with sub-planckian curvature}} (\bibinfo {year} {2024}),\ \Eprint {https://arxiv.org/abs/2402.15085} {arXiv:2402.15085 [id='gr-qc']} \BibitemShut {NoStop}%
\bibitem [{\citenamefont {Ghosh}\ \emph {et~al.}(2023)\citenamefont {Ghosh}, \citenamefont {Rahman},\ and\ \citenamefont {Mishra}}]{Ghosh_2023}%
  \BibitemOpen
  \bibfield  {author} {\bibinfo {author} {\bibfnamefont {R.}~\bibnamefont {Ghosh}}, \bibinfo {author} {\bibfnamefont {M.}~\bibnamefont {Rahman}},\ and\ \bibinfo {author} {\bibfnamefont {A.~K.}\ \bibnamefont {Mishra}},\ }\bibfield  {title} {\bibinfo {title} {Regularized stable kerr black hole: cosmic censorships, shadow and quasi-normal modes},\ }\bibfield  {journal} {\bibinfo  {journal} {The European Physical Journal C}\ }\textbf {\bibinfo {volume} {83}},\ \href {https://doi.org/10.1140/epjc/s10052-023-11252-0} {10.1140/epjc/s10052-023-11252-0} (\bibinfo {year} {2023})\BibitemShut {NoStop}%
\bibitem [{\citenamefont {{Pedrotti}}\ and\ \citenamefont {{Vagnozzi}}(2024)}]{2024arXiv240407589P}%
  \BibitemOpen
  \bibfield  {author} {\bibinfo {author} {\bibfnamefont {D.}~\bibnamefont {{Pedrotti}}}\ and\ \bibinfo {author} {\bibfnamefont {S.}~\bibnamefont {{Vagnozzi}}},\ }\bibfield  {title} {\bibinfo {title} {{See the lightning, hear the thunder: quasinormal modes-shadow correspondence for rotating regular black holes}},\ }\href {https://doi.org/10.48550/arXiv.2404.07589} {\bibfield  {journal} {\bibinfo  {journal} {arXiv e-prints}\ ,\ \bibinfo {eid} {arXiv:2404.07589}} (\bibinfo {year} {2024})},\ \Eprint {https://arxiv.org/abs/2404.07589} {arXiv:2404.07589 [gr-qc]} \BibitemShut {NoStop}%
\bibitem [{\citenamefont {{Markov}}\ and\ \citenamefont {{Mukhanov}}(1985)}]{1985NCimB..86...97M}%
  \BibitemOpen
  \bibfield  {author} {\bibinfo {author} {\bibfnamefont {M.~A.}\ \bibnamefont {{Markov}}}\ and\ \bibinfo {author} {\bibfnamefont {V.~F.}\ \bibnamefont {{Mukhanov}}},\ }\bibfield  {title} {\bibinfo {title} {{De Sitter-like initial state of the universe as a result of asymptotical disappearance of gravitational interactions of matter.}},\ }\href {https://doi.org/10.1007/BF02732276} {\bibfield  {journal} {\bibinfo  {journal} {Nuovo Cimento B Serie}\ }\textbf {\bibinfo {volume} {86B}},\ \bibinfo {pages} {97} (\bibinfo {year} {1985})}\BibitemShut {NoStop}%
\bibitem [{\citenamefont {Reuter}(1998)}]{PhysRevD.57.971}%
  \BibitemOpen
  \bibfield  {author} {\bibinfo {author} {\bibfnamefont {M.}~\bibnamefont {Reuter}},\ }\bibfield  {title} {\bibinfo {title} {Nonperturbative evolution equation for quantum gravity},\ }\href {https://doi.org/10.1103/PhysRevD.57.971} {\bibfield  {journal} {\bibinfo  {journal} {Phys. Rev. D}\ }\textbf {\bibinfo {volume} {57}},\ \bibinfo {pages} {971} (\bibinfo {year} {1998})}\BibitemShut {NoStop}%
\bibitem [{\citenamefont {Bonanno}\ \emph {et~al.}(2020)\citenamefont {Bonanno}, \citenamefont {Casadio},\ and\ \citenamefont {Platania}}]{Bonanno_2020}%
  \BibitemOpen
  \bibfield  {author} {\bibinfo {author} {\bibfnamefont {A.}~\bibnamefont {Bonanno}}, \bibinfo {author} {\bibfnamefont {R.}~\bibnamefont {Casadio}},\ and\ \bibinfo {author} {\bibfnamefont {A.}~\bibnamefont {Platania}},\ }\bibfield  {title} {\bibinfo {title} {Gravitational antiscreening in stellar interiors},\ }\href {https://doi.org/10.1088/1475-7516/2020/01/022} {\bibfield  {journal} {\bibinfo  {journal} {Journal of Cosmology and Astroparticle Physics}\ }\textbf {\bibinfo {volume} {2020}}\bibinfo  {number} { (01)},\ \bibinfo {pages} {022}}\BibitemShut {NoStop}%
\bibitem [{\citenamefont {Bonanno}\ \emph {et~al.}(2022)\citenamefont {Bonanno}, \citenamefont {Denz}, \citenamefont {Pawlowski},\ and\ \citenamefont {Reichert}}]{10.21468/SciPostPhys.12.1.001}%
  \BibitemOpen
\bibfield  {number} {  }\bibfield  {author} {\bibinfo {author} {\bibfnamefont {A.}~\bibnamefont {Bonanno}}, \bibinfo {author} {\bibfnamefont {T.}~\bibnamefont {Denz}}, \bibinfo {author} {\bibfnamefont {J.~M.}\ \bibnamefont {Pawlowski}},\ and\ \bibinfo {author} {\bibfnamefont {M.}~\bibnamefont {Reichert}},\ }\bibfield  {title} {\bibinfo {title} {{Reconstructing the graviton}},\ }\href {https://doi.org/10.21468/SciPostPhys.12.1.001} {\bibfield  {journal} {\bibinfo  {journal} {SciPost Phys.}\ }\textbf {\bibinfo {volume} {12}},\ \bibinfo {pages} {001} (\bibinfo {year} {2022})}\BibitemShut {NoStop}%
\bibitem [{\citenamefont {Bonanno}\ and\ \citenamefont {Reuter}(2000{\natexlab{b}})}]{PhysRevD.62.043008}%
  \BibitemOpen
  \bibfield  {author} {\bibinfo {author} {\bibfnamefont {A.}~\bibnamefont {Bonanno}}\ and\ \bibinfo {author} {\bibfnamefont {M.}~\bibnamefont {Reuter}},\ }\bibfield  {title} {\bibinfo {title} {Renormalization group improved black hole spacetimes},\ }\href {https://doi.org/10.1103/PhysRevD.62.043008} {\bibfield  {journal} {\bibinfo  {journal} {Phys. Rev. D}\ }\textbf {\bibinfo {volume} {62}},\ \bibinfo {pages} {043008} (\bibinfo {year} {2000}{\natexlab{b}})}\BibitemShut {NoStop}%
\bibitem [{\citenamefont {Brill}\ and\ \citenamefont {Wheeler}(1957)}]{RevModPhys.29.465}%
  \BibitemOpen
  \bibfield  {author} {\bibinfo {author} {\bibfnamefont {D.~R.}\ \bibnamefont {Brill}}\ and\ \bibinfo {author} {\bibfnamefont {J.~A.}\ \bibnamefont {Wheeler}},\ }\bibfield  {title} {\bibinfo {title} {Interaction of neutrinos and gravitational fields},\ }\href {https://doi.org/10.1103/RevModPhys.29.465} {\bibfield  {journal} {\bibinfo  {journal} {Rev. Mod. Phys.}\ }\textbf {\bibinfo {volume} {29}},\ \bibinfo {pages} {465} (\bibinfo {year} {1957})}\BibitemShut {NoStop}%
\bibitem [{\citenamefont {Konoplya}\ and\ \citenamefont {Zhidenko}(2011)}]{Konoplya_2011}%
  \BibitemOpen
  \bibfield  {author} {\bibinfo {author} {\bibfnamefont {R.~A.}\ \bibnamefont {Konoplya}}\ and\ \bibinfo {author} {\bibfnamefont {A.}~\bibnamefont {Zhidenko}},\ }\bibfield  {title} {\bibinfo {title} {Quasinormal modes of black holes: From astrophysics to string theory},\ }\href {https://doi.org/10.1103/revmodphys.83.793} {\bibfield  {journal} {\bibinfo  {journal} {Reviews of Modern Physics}\ }\textbf {\bibinfo {volume} {83}},\ \bibinfo {pages} {793–836} (\bibinfo {year} {2011})}\BibitemShut {NoStop}%
\bibitem [{\citenamefont {Arbey}\ \emph {et~al.}(2021)\citenamefont {Arbey}, \citenamefont {Auffinger}, \citenamefont {Geiller}, \citenamefont {Livine},\ and\ \citenamefont {Sartini}}]{Arbey_2021}%
  \BibitemOpen
  \bibfield  {author} {\bibinfo {author} {\bibfnamefont {A.}~\bibnamefont {Arbey}}, \bibinfo {author} {\bibfnamefont {J.}~\bibnamefont {Auffinger}}, \bibinfo {author} {\bibfnamefont {M.}~\bibnamefont {Geiller}}, \bibinfo {author} {\bibfnamefont {E.~R.}\ \bibnamefont {Livine}},\ and\ \bibinfo {author} {\bibfnamefont {F.}~\bibnamefont {Sartini}},\ }\bibfield  {title} {\bibinfo {title} {Hawking radiation by spherically-symmetric static black holes for all spins: Teukolsky equations and potentials},\ }\bibfield  {journal} {\bibinfo  {journal} {Physical Review D}\ }\textbf {\bibinfo {volume} {103}},\ \href {https://doi.org/10.1103/physrevd.103.104010} {10.1103/physrevd.103.104010} (\bibinfo {year} {2021})\BibitemShut {NoStop}%
\bibitem [{\citenamefont {Zinhailo}(2019)}]{Zinhailo_2019}%
  \BibitemOpen
  \bibfield  {author} {\bibinfo {author} {\bibfnamefont {A.~F.}\ \bibnamefont {Zinhailo}},\ }\bibfield  {title} {\bibinfo {title} {Quasinormal modes of dirac field in the einstein–dilaton–gauss–bonnet and einstein–weyl gravities},\ }\bibfield  {journal} {\bibinfo  {journal} {The European Physical Journal C}\ }\textbf {\bibinfo {volume} {79}},\ \href {https://doi.org/10.1140/epjc/s10052-019-7425-9} {10.1140/epjc/s10052-019-7425-9} (\bibinfo {year} {2019})\BibitemShut {NoStop}%
\bibitem [{\citenamefont {Boyd}(2013)}]{boyd2013chebyshev}%
  \BibitemOpen
  \bibfield  {author} {\bibinfo {author} {\bibfnamefont {J.}~\bibnamefont {Boyd}},\ }\href {https://books.google.com/books?id=b4TCAgAAQBAJ} {\emph {\bibinfo {title} {Chebyshev and Fourier Spectral Methods: Second Revised Edition}}},\ Dover Books on Mathematics\ (\bibinfo  {publisher} {Dover Publications},\ \bibinfo {year} {2013})\BibitemShut {NoStop}%
\bibitem [{\citenamefont {{Jansen}}(2017)}]{2017EPJP..132..546J}%
  \BibitemOpen
  \bibfield  {author} {\bibinfo {author} {\bibfnamefont {A.}~\bibnamefont {{Jansen}}},\ }\bibfield  {title} {\bibinfo {title} {{Overdamped modes in Schwarzschild-de Sitter and a Mathematica package for the numerical computation of quasinormal modes}},\ }\href {https://doi.org/10.1140/epjp/i2017-11825-9} {\bibfield  {journal} {\bibinfo  {journal} {European Physical Journal Plus}\ }\textbf {\bibinfo {volume} {132}},\ \bibinfo {eid} {546} (\bibinfo {year} {2017})},\ \Eprint {https://arxiv.org/abs/1709.09178} {arXiv:1709.09178 [gr-qc]} \BibitemShut {NoStop}%
\bibitem [{\citenamefont {Berti}\ and\ \citenamefont {Kokkotas}(2003)}]{PhysRevD.68.044027}%
  \BibitemOpen
  \bibfield  {author} {\bibinfo {author} {\bibfnamefont {E.}~\bibnamefont {Berti}}\ and\ \bibinfo {author} {\bibfnamefont {K.~D.}\ \bibnamefont {Kokkotas}},\ }\bibfield  {title} {\bibinfo {title} {Asymptotic quasinormal modes of reissner-nordstr\"om and kerr black holes},\ }\href {https://doi.org/10.1103/PhysRevD.68.044027} {\bibfield  {journal} {\bibinfo  {journal} {Phys. Rev. D}\ }\textbf {\bibinfo {volume} {68}},\ \bibinfo {pages} {044027} (\bibinfo {year} {2003})}\BibitemShut {NoStop}%
\bibitem [{\citenamefont {Jing}(2005)}]{Jiliang2005}%
  \BibitemOpen
  \bibfield  {author} {\bibinfo {author} {\bibfnamefont {J.}~\bibnamefont {Jing}},\ }\bibfield  {title} {\bibinfo {title} {Neutrino quasinormal modes of the reissner-nordström black hole},\ }\href {https://doi.org/10.1088/1126-6708/2005/12/005} {\bibfield  {journal} {\bibinfo  {journal} {Journal of High Energy Physics}\ }\textbf {\bibinfo {volume} {2005}},\ \bibinfo {pages} {005} (\bibinfo {year} {2005})}\BibitemShut {NoStop}%
\bibitem [{\citenamefont {Richartz}(2016)}]{PhysRevD.93.064062}%
  \BibitemOpen
  \bibfield  {author} {\bibinfo {author} {\bibfnamefont {M.}~\bibnamefont {Richartz}},\ }\bibfield  {title} {\bibinfo {title} {Quasinormal modes of extremal black holes},\ }\href {https://doi.org/10.1103/PhysRevD.93.064062} {\bibfield  {journal} {\bibinfo  {journal} {Phys. Rev. D}\ }\textbf {\bibinfo {volume} {93}},\ \bibinfo {pages} {064062} (\bibinfo {year} {2016})}\BibitemShut {NoStop}%
\bibitem [{\citenamefont {Konoplya}\ \emph {et~al.}(2022)\citenamefont {Konoplya}, \citenamefont {Zinhailo}, \citenamefont {Kunz}, \citenamefont {Stuchlík},\ and\ \citenamefont {Zhidenko}}]{Konoplya_2022}%
  \BibitemOpen
  \bibfield  {author} {\bibinfo {author} {\bibfnamefont {R.}~\bibnamefont {Konoplya}}, \bibinfo {author} {\bibfnamefont {A.}~\bibnamefont {Zinhailo}}, \bibinfo {author} {\bibfnamefont {J.}~\bibnamefont {Kunz}}, \bibinfo {author} {\bibfnamefont {Z.}~\bibnamefont {Stuchlík}},\ and\ \bibinfo {author} {\bibfnamefont {A.}~\bibnamefont {Zhidenko}},\ }\bibfield  {title} {\bibinfo {title} {Quasinormal ringing of regular black holes in asymptotically safe gravity: the importance of overtones},\ }\href {https://doi.org/10.1088/1475-7516/2022/10/091} {\bibfield  {journal} {\bibinfo  {journal} {Journal of Cosmology and Astroparticle Physics}\ }\textbf {\bibinfo {volume} {2022}}\bibinfo  {number} { (10)},\ \bibinfo {pages} {091}}\BibitemShut {NoStop}%
\bibitem [{\citenamefont {{Fu}}\ \emph {et~al.}(2024)\citenamefont {{Fu}}, \citenamefont {{Zhang}}, \citenamefont {{Liu}}, \citenamefont {{Kuang}},\ and\ \citenamefont {{Wu}}}]{2024PhRvD.109b6010F}%
  \BibitemOpen
\bibfield  {number} {  }\bibfield  {author} {\bibinfo {author} {\bibfnamefont {G.}~\bibnamefont {{Fu}}}, \bibinfo {author} {\bibfnamefont {D.}~\bibnamefont {{Zhang}}}, \bibinfo {author} {\bibfnamefont {P.}~\bibnamefont {{Liu}}}, \bibinfo {author} {\bibfnamefont {X.-M.}\ \bibnamefont {{Kuang}}},\ and\ \bibinfo {author} {\bibfnamefont {J.-P.}\ \bibnamefont {{Wu}}},\ }\bibfield  {title} {\bibinfo {title} {{Peculiar properties in quasinormal spectra from loop quantum gravity effect}},\ }\href {https://doi.org/10.1103/PhysRevD.109.026010} {\bibfield  {journal} {\bibinfo  {journal} {\prd}\ }\textbf {\bibinfo {volume} {109}},\ \bibinfo {eid} {026010} (\bibinfo {year} {2024})},\ \Eprint {https://arxiv.org/abs/2301.08421} {arXiv:2301.08421 [gr-qc]} \BibitemShut {NoStop}%
\bibitem [{\citenamefont {Moreira}\ \emph {et~al.}(2023)\citenamefont {Moreira}, \citenamefont {Junior}, \citenamefont {Crispino},\ and\ \citenamefont {Herdeiro}}]{Moreira_2023}%
  \BibitemOpen
  \bibfield  {author} {\bibinfo {author} {\bibfnamefont {Z.~S.}\ \bibnamefont {Moreira}}, \bibinfo {author} {\bibfnamefont {H.~C.~L.}\ \bibnamefont {Junior}}, \bibinfo {author} {\bibfnamefont {L.~C.}\ \bibnamefont {Crispino}},\ and\ \bibinfo {author} {\bibfnamefont {C.~A.}\ \bibnamefont {Herdeiro}},\ }\bibfield  {title} {\bibinfo {title} {Quasinormal modes of a holonomy corrected schwarzschild black hole},\ }\bibfield  {journal} {\bibinfo  {journal} {Physical Review D}\ }\textbf {\bibinfo {volume} {107}},\ \href {https://doi.org/10.1103/physrevd.107.104016} {10.1103/physrevd.107.104016} (\bibinfo {year} {2023})\BibitemShut {NoStop}%
\bibitem [{\citenamefont {{Gong}}\ \emph {et~al.}(2023)\citenamefont {{Gong}}, \citenamefont {{Li}}, \citenamefont {{Zhang}}, \citenamefont {{Fu}},\ and\ \citenamefont {{Wu}}}]{2023arXiv231217639G}%
  \BibitemOpen
  \bibfield  {author} {\bibinfo {author} {\bibfnamefont {H.}~\bibnamefont {{Gong}}}, \bibinfo {author} {\bibfnamefont {S.}~\bibnamefont {{Li}}}, \bibinfo {author} {\bibfnamefont {D.}~\bibnamefont {{Zhang}}}, \bibinfo {author} {\bibfnamefont {G.}~\bibnamefont {{Fu}}},\ and\ \bibinfo {author} {\bibfnamefont {J.-P.}\ \bibnamefont {{Wu}}},\ }\bibfield  {title} {\bibinfo {title} {{Quasinormal modes of quantum-corrected black holes}},\ }\href {https://doi.org/10.48550/arXiv.2312.17639} {\bibfield  {journal} {\bibinfo  {journal} {arXiv e-prints}\ ,\ \bibinfo {eid} {arXiv:2312.17639}} (\bibinfo {year} {2023})},\ \Eprint {https://arxiv.org/abs/2312.17639} {arXiv:2312.17639 [gr-qc]} \BibitemShut {NoStop}%
\bibitem [{\citenamefont {{Gingrich}}(2024)}]{2024arXiv240404447G}%
  \BibitemOpen
  \bibfield  {author} {\bibinfo {author} {\bibfnamefont {D.~M.}\ \bibnamefont {{Gingrich}}},\ }\bibfield  {title} {\bibinfo {title} {{Quasinormal modes of a nonsingular spherically symmetric black hole effective model with holonomy corrections}},\ }\href {https://doi.org/10.48550/arXiv.2404.04447} {\bibfield  {journal} {\bibinfo  {journal} {arXiv e-prints}\ ,\ \bibinfo {eid} {arXiv:2404.04447}} (\bibinfo {year} {2024})},\ \Eprint {https://arxiv.org/abs/2404.04447} {arXiv:2404.04447 [gr-qc]} \BibitemShut {NoStop}%
\bibitem [{\citenamefont {{ D. {Psaltis} and \textit{et al}}}(2020)}]{PhysRevLett.125.141104}%
  \BibitemOpen
  \bibfield  {author} {\bibinfo {author} {\bibnamefont {{ D. {Psaltis} and \textit{et al}}}} (\bibinfo {collaboration} {Event Horizon Telescope Collaboration}),\ }\bibfield  {title} {\bibinfo {title} {Gravitational test beyond the first post-newtonian order with the shadow of the m87 black hole},\ }\href {https://doi.org/10.1103/PhysRevLett.125.141104} {\bibfield  {journal} {\bibinfo  {journal} {Phys. Rev. Lett.}\ }\textbf {\bibinfo {volume} {125}},\ \bibinfo {pages} {141104} (\bibinfo {year} {2020})}\BibitemShut {NoStop}%
\bibitem [{\citenamefont {Vagnozzi}\ \emph {et~al.}(2023)\citenamefont {Vagnozzi}, \citenamefont {Roy}, \citenamefont {Tsai}, \citenamefont {Visinelli}, \citenamefont {Afrin}, \citenamefont {Allahyari}, \citenamefont {Bambhaniya}, \citenamefont {Dey}, \citenamefont {Ghosh}, \citenamefont {Joshi}, \citenamefont {Jusufi}, \citenamefont {Khodadi}, \citenamefont {Walia}, \citenamefont {Övgün},\ and\ \citenamefont {Bambi}}]{Vagnozzi_2023}%
  \BibitemOpen
  \bibfield  {author} {\bibinfo {author} {\bibfnamefont {S.}~\bibnamefont {Vagnozzi}}, \bibinfo {author} {\bibfnamefont {R.}~\bibnamefont {Roy}}, \bibinfo {author} {\bibfnamefont {Y.-D.}\ \bibnamefont {Tsai}}, \bibinfo {author} {\bibfnamefont {L.}~\bibnamefont {Visinelli}}, \bibinfo {author} {\bibfnamefont {M.}~\bibnamefont {Afrin}}, \bibinfo {author} {\bibfnamefont {A.}~\bibnamefont {Allahyari}}, \bibinfo {author} {\bibfnamefont {P.}~\bibnamefont {Bambhaniya}}, \bibinfo {author} {\bibfnamefont {D.}~\bibnamefont {Dey}}, \bibinfo {author} {\bibfnamefont {S.~G.}\ \bibnamefont {Ghosh}}, \bibinfo {author} {\bibfnamefont {P.~S.}\ \bibnamefont {Joshi}}, \bibinfo {author} {\bibfnamefont {K.}~\bibnamefont {Jusufi}}, \bibinfo {author} {\bibfnamefont {M.}~\bibnamefont {Khodadi}}, \bibinfo {author} {\bibfnamefont {R.~K.}\ \bibnamefont {Walia}}, \bibinfo {author} {\bibfnamefont {A.}~\bibnamefont {Övgün}},\ and\ \bibinfo {author} {\bibfnamefont {C.}~\bibnamefont {Bambi}},\ }\bibfield  {title} {\bibinfo {title}
  {Horizon-scale tests of gravity theories and fundamental physics from the event horizon telescope image of sagittarius a},\ }\href {https://doi.org/10.1088/1361-6382/acd97b} {\bibfield  {journal} {\bibinfo  {journal} {Classical and Quantum Gravity}\ }\textbf {\bibinfo {volume} {40}},\ \bibinfo {pages} {165007} (\bibinfo {year} {2023})}\BibitemShut {NoStop}%
\bibitem [{\citenamefont {Cardoso}\ \emph {et~al.}(2009)\citenamefont {Cardoso}, \citenamefont {Miranda}, \citenamefont {Berti}, \citenamefont {Witek},\ and\ \citenamefont {Zanchin}}]{PhysRevD.79.064016}%
  \BibitemOpen
  \bibfield  {author} {\bibinfo {author} {\bibfnamefont {V.}~\bibnamefont {Cardoso}}, \bibinfo {author} {\bibfnamefont {A.~S.}\ \bibnamefont {Miranda}}, \bibinfo {author} {\bibfnamefont {E.}~\bibnamefont {Berti}}, \bibinfo {author} {\bibfnamefont {H.}~\bibnamefont {Witek}},\ and\ \bibinfo {author} {\bibfnamefont {V.~T.}\ \bibnamefont {Zanchin}},\ }\bibfield  {title} {\bibinfo {title} {Geodesic stability, lyapunov exponents, and quasinormal modes},\ }\href {https://doi.org/10.1103/PhysRevD.79.064016} {\bibfield  {journal} {\bibinfo  {journal} {Phys. Rev. D}\ }\textbf {\bibinfo {volume} {79}},\ \bibinfo {pages} {064016} (\bibinfo {year} {2009})}\BibitemShut {NoStop}%
\bibitem [{\citenamefont {Jusufi}(2020)}]{Jusufi_2020}%
  \BibitemOpen
  \bibfield  {author} {\bibinfo {author} {\bibfnamefont {K.}~\bibnamefont {Jusufi}},\ }\bibfield  {title} {\bibinfo {title} {Connection between the shadow radius and quasinormal modes in rotating spacetimes},\ }\bibfield  {journal} {\bibinfo  {journal} {Physical Review D}\ }\textbf {\bibinfo {volume} {101}},\ \href {https://doi.org/10.1103/physrevd.101.124063} {10.1103/physrevd.101.124063} (\bibinfo {year} {2020})\BibitemShut {NoStop}%
\bibitem [{\citenamefont {Yang}(2021)}]{Yang_2021}%
  \BibitemOpen
  \bibfield  {author} {\bibinfo {author} {\bibfnamefont {H.}~\bibnamefont {Yang}},\ }\bibfield  {title} {\bibinfo {title} {Relating black hole shadow to quasinormal modes for rotating black holes},\ }\bibfield  {journal} {\bibinfo  {journal} {Physical Review D}\ }\textbf {\bibinfo {volume} {103}},\ \href {https://doi.org/10.1103/physrevd.103.084010} {10.1103/physrevd.103.084010} (\bibinfo {year} {2021})\BibitemShut {NoStop}%
\bibitem [{\citenamefont {Konoplya}\ \emph {et~al.}(2019)\citenamefont {Konoplya}, \citenamefont {Zhidenko},\ and\ \citenamefont {Zinhailo}}]{Konoplya_2019}%
  \BibitemOpen
  \bibfield  {author} {\bibinfo {author} {\bibfnamefont {R.~A.}\ \bibnamefont {Konoplya}}, \bibinfo {author} {\bibfnamefont {A.}~\bibnamefont {Zhidenko}},\ and\ \bibinfo {author} {\bibfnamefont {A.~F.}\ \bibnamefont {Zinhailo}},\ }\bibfield  {title} {\bibinfo {title} {Higher order wkb formula for quasinormal modes and grey-body factors: recipes for quick and accurate calculations},\ }\href {https://doi.org/10.1088/1361-6382/ab2e25} {\bibfield  {journal} {\bibinfo  {journal} {Classical and Quantum Gravity}\ }\textbf {\bibinfo {volume} {36}},\ \bibinfo {pages} {155002} (\bibinfo {year} {2019})}\BibitemShut {NoStop}%
\bibitem [{\citenamefont {Konoplya}(2003)}]{Konoplya_2003}%
  \BibitemOpen
  \bibfield  {author} {\bibinfo {author} {\bibfnamefont {R.~A.}\ \bibnamefont {Konoplya}},\ }\bibfield  {title} {\bibinfo {title} {Quasinormal behavior of the d dimensional schwarzschild black hole and the higher order wkb approach},\ }\bibfield  {journal} {\bibinfo  {journal} {Physical Review D}\ }\textbf {\bibinfo {volume} {68}},\ \href {https://doi.org/10.1103/physrevd.68.024018} {10.1103/physrevd.68.024018} (\bibinfo {year} {2003})\BibitemShut {NoStop}%
\bibitem [{\citenamefont {Visser}(1999)}]{PhysRevA.59.427}%
  \BibitemOpen
  \bibfield  {author} {\bibinfo {author} {\bibfnamefont {M.}~\bibnamefont {Visser}},\ }\bibfield  {title} {\bibinfo {title} {Some general bounds for one-dimensional scattering},\ }\href {https://doi.org/10.1103/PhysRevA.59.427} {\bibfield  {journal} {\bibinfo  {journal} {Phys. Rev. A}\ }\textbf {\bibinfo {volume} {59}},\ \bibinfo {pages} {427} (\bibinfo {year} {1999})}\BibitemShut {NoStop}%
\bibitem [{\citenamefont {{Boonserm}}\ and\ \citenamefont {{Visser}}(2008)}]{2008AnPhy.323.2779B}%
  \BibitemOpen
  \bibfield  {author} {\bibinfo {author} {\bibfnamefont {P.}~\bibnamefont {{Boonserm}}}\ and\ \bibinfo {author} {\bibfnamefont {M.}~\bibnamefont {{Visser}}},\ }\bibfield  {title} {\bibinfo {title} {{Bounding the Bogoliubov coefficients}},\ }\href {https://doi.org/10.1016/j.aop.2008.02.002} {\bibfield  {journal} {\bibinfo  {journal} {Annals of Physics}\ }\textbf {\bibinfo {volume} {323}},\ \bibinfo {pages} {2779} (\bibinfo {year} {2008})},\ \Eprint {https://arxiv.org/abs/0801.0610} {arXiv:0801.0610 [quant-ph]} \BibitemShut {NoStop}%
\end{thebibliography}%

\end{document}